\PassOptionsToPackage{table}{xcolor}
\PassOptionsToPackage{capitalize}{cleveref}
\documentclass{spml_template}

\usepackage[utf8]{inputenc} 
\usepackage[T1]{fontenc}    
\usepackage{hyperref}       
\usepackage{url}            
\usepackage{booktabs}       
\usepackage{amsfonts}       
\usepackage{nicefrac}       
\usepackage{microtype}      
\usepackage[table]{xcolor}         

\usepackage{graphicx}
\usepackage{amsmath}
\usepackage{amssymb, bm}
\usepackage{mathtools}
\usepackage{amsthm}
\usepackage{algorithm}
\usepackage{algpseudocode}
\usepackage{multirow}
\usepackage{siunitx}
\usepackage{makecell}
\usepackage{wrapfig}
\usepackage{subcaption}
\usepackage{xspace}
\usepackage{soul}
\usepackage{pifont}
\usepackage{float}
\usepackage{chemformula}
\usepackage[varqu]{zi4}     
\usepackage[version=4]{mhchem} 
\usepackage[capitalize, noabbrev, nameinlink]{cleveref}
\usepackage{enumitem}
\usepackage{fontawesome}

\providecommand{\appendixcontentsstart}{}

\newcommand{\fkhyperparamwidth}{1.0\linewidth}
\newif\iffkhyperparamwide 
\newcommand{\concepttopskip}{-3pt}
\newcommand{\conceptbottomskip}{-5pt}
\newcommand{\fktabletopskip}{-10pt}
\newcommand{\fktablebottomskip}{-3pt}
\newcommand{\fktablewidth}{0.46\textwidth}
\newcommand{\fktablelines}{}
\newcommand{\scalingfigwidth}{0.48\textwidth}
\newcommand{\scalingfigsep}{0.8em}
\newcommand{\odacplotheight}{2.5cm}
\newcommand{\propertytablewidth}{0.46\linewidth}
\newcommand{\bwdbtablestretch}{1.05}
\newcommand{\bwdbtabcolsep}{4pt}
\newcommand{\bwdbtablebottomskip}{-13pt}
\newcommand{\successtablewidth}{0.42\textwidth}
\newcommand{\successtabletopskip}{-1.0em}
\newcommand{\odacviswidth}{0.43\linewidth}
\newcommand{\odacvisscale}{1}

\definecolor{rowhl}{HTML}{EEEAFE}
\definecolor{myblue}{RGB}{0, 0, 255} 
\definecolor{mydarkblue}{RGB}{0, 0, 139} 
\definecolor{seethis}{RGB}{255, 127, 127} 
\definecolor{Gray}{gray}{0.85}
\definecolor{lightorange}{RGB}{252,245,245}
\definecolor{red}{RGB}{252,245,245}

\definecolor{PrimaryBlue}{HTML}{5C41BF}
\definecolor{SecondaryPurple}{HTML}{7C69BF}
\definecolor{SoftLavender}{HTML}{E7DFF2}
\definecolor{MutedOlive}{HTML}{A69D8D}
\definecolor{InkBlack}{HTML}{0D0D0D}

\definecolor{MyBlue2}{HTML}{5384EC}
\definecolor{MyYellow2}{HTML}{F5B025}
\definecolor{MyRed2}{HTML}{BF0449}

\newcommand{\ev}{\text{eV}}
\newcommand{\angstrom}{\text{\normalfont\AA}}

\newcommand{\eva}{\ev/\angstrom}

\definecolor{ourscolor}{HTML}{E9E5F3}
\newcommand{\ours}{\textsc{AtomMOF}}

\theoremstyle{plain}

\theoremstyle{definition}

\theoremstyle{remark}

\title{\ours{}: All-Atom Flow Matching for MOF-Adsorbate Structure Prediction}

\renewcommand{\successtablewidth}{0.38\textwidth}
\renewcommand{\concepttopskip}{-8.5pt}
\renewcommand{\conceptbottomskip}{-12.5pt}
\fkhyperparamwidetrue
\renewcommand{\fktabletopskip}{-16pt}
\renewcommand{\fktablebottomskip}{-6pt}
\renewcommand{\scalingfigwidth}{0.44\textwidth}
\renewcommand{\successtabletopskip}{-1.6em}
\renewcommand{\odacvisscale}{0.88}

\author[1]{Nayoung Kim}
\author[1]{Honghui Kim}
\author[1]{Sihyun Yu}
\author[1]{Minkyu Kim}
\author[1]{Seongsu Kim}
\author[1]{Sungsoo Ahn}

\affiliation[1]{Korea Advanced Institute of Science and Technology (KAIST)}

\metadata[Code]{\href{https://github.com/nayoung10/AtomMOF}{\texttt{https://github.com/nayoung10/AtomMOF}}}
\correspondence{\email{nayoungkim@kaist.ac.kr}, \email{sungsoo.ahn@kaist.ac.kr}}

\abstract{Metal-organic frameworks (MOFs) are promising materials for applications such as direct air capture, where performance depends on how adsorbates bind within the framework. Accurately predicting MOF-adsorbate configurations is therefore important for screening candidate materials. However, traditional approaches are computationally expensive and require a known host structure, while existing generative models rely on rigid-body assumptions and do not explicitly model adsorbates. We introduce \ours{}, a scalable all-atom flow-matching model that jointly predicts MOF-adsorbate structures from discrete building blocks and adsorbates. Built on a Diffusion Transformer with a building block-based pairwise attention bias, \ours{} operates in an unconstrained all-atom space and exhibits clear scaling behavior. To improve the structural validity of flexible all-atom models, we also propose Feynman-Kac (FK) steering with machine-learned interatomic potentials (MLIPs). On the BW dataset, \ours{} achieves a 85.61\% relative increase in match rate and a 34.32\% relative reduction in RMSD compared with prior work. MLIP-guided steering further improves validity by 28.7\% and reduces formation energy error by 86.5\%. On ODAC25, \ours{} generates adsorption configurations faster than GCMC and, when combined with MLIP relaxation, identifies lower-energy configurations than those in the reference dataset.
}

\begin{document}
\maketitle

\section{Introduction} \label{intro}

\begin{wrapfigure}{r}{0.48\linewidth}
    \centering
    \vspace{\concepttopskip}
    \includegraphics[width=\linewidth]{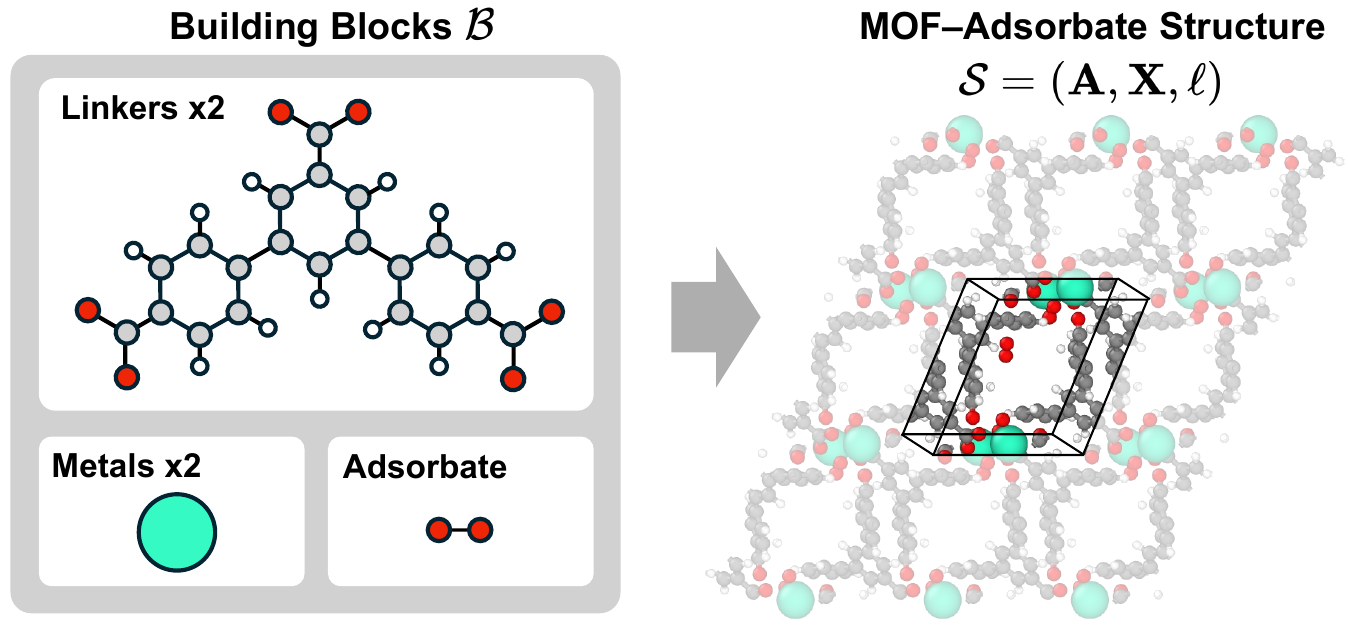}
    \caption{\textbf{MOF-adsorbate structure prediction.} Given building blocks $\mathcal{B}$, including linkers, metal nodes, and adsorbate, the goal is to predict the equilibrium structure $\mathcal{S} = (\mathbf{A}, \mathbf{X}, \bm{\ell})$ with atomic numbers $\mathbf{A}$, coordinates $\mathbf{X}$, and lattice $\bm{\ell}$.}
    \label{fig:concept}
    \vspace{\conceptbottomskip}
\end{wrapfigure}


Metal-organic frameworks (MOFs) are porous crystalline materials with large surface areas and highly tunable chemistries. These features make them promising for applications such as gas separation \citep{qian2020mof}, catalysis \citep{lee2009metal}, and sensing \citep{kreno2012metal}. One important application is direct air capture (DAC), which requires selectively adsorbing \ce{CO2} from ambient air. This task is inherently challenging because atmospheric \ce{CO2} is extremely dilute, while competing species such as \ce{H2O} can occupy adsorption sites \citep{bose2024challenges}. Because adsorption performance depends on where adsorbates bind within the framework, accurately predicting MOF-adsorbate configurations is essential for discovering effective carbon capture materials.

Traditional approaches for predicting MOF-adsorbate binding configurations rely on search-based, iterative workflows. They first generate candidate adsorbate placements using grand canonical Monte Carlo~\citep[GCMC;][]{adams1975grand}, heuristics, or random sampling~\citep{lan2023adsorbml}, and then relax each candidate to a local energy minimum via iterative energy optimization. Although effective, these workflows are computationally expensive for MOFs, which often have hundreds of atoms per unit cell. Moreover, they require a known 3D host structure, limiting screening to existing MOF databases despite the vast space of hypothetical materials.

To overcome these challenges, deep generative models have emerged as a promising alternative. However, existing models have two key limitations. First, they often operate at a coarse-grained resolution with rigid-body assumptions~\citep{kim2025mofflow, jiao2025mof, pan2026enhancing} or fixed bond lengths and angles \citep{kim2025flexible}. Although this simplifies the modeling task, it fails to capture variations in bond lengths, bond angles, and coordination geometries arising from local chemical environments and MOF-adsorbate interactions. Second, these methods generate the MOF structure in isolation, neglecting the adsorbates that ultimately govern performance in applications such as DAC. Because MOF-adsorbate configurations determine adsorption energies, directly predicting them accelerates materials screening.

\textbf{Contribution.} We introduce \ours{}, a scalable flow-matching model for all-atom MOF-adsorbate structure prediction. Given only discrete building block identities and adsorbate types, \ours{} directly predicts low-energy MOF-adsorbate configurations (\Cref{fig:concept}). Built on a unified Diffusion Transformer~\citep[DiT;][]{peebles2023scalable}, \ours{} augments attention with a pairwise bias that encodes local geometry and connectivity within each building block. This architecture supports fully unconstrained all-atom generation while jointly modeling frameworks and adsorbates.

This formulation makes structure prediction more useful for MOF screening. Unlike existing MOF generative models that are structurally constrained, \ours{} predicts all atoms in an unconstrained coordinate space and can therefore capture fine-grained variations in bond lengths, bond angles, and coordination environments. Moreover, because it predicts the MOF and adsorbate jointly, the resulting structures can be evaluated with application-relevant properties such as adsorption energy.

However, this flexibility introduces an additional challenge: unconstrained all-atom models can produce local geometric distortions, such as deformed aromatic rings~\citep{jiang2025posex}. To address this issue, we introduce Feynman-Kac steering with machine-learning interatomic potentials (MLIPs), a simple inference-time strategy that improves structural quality and stability without retraining. Unlike prior steering approaches that rely on hand-designed energy functions~\citep{wohlwend2025boltz} or expensive multi-stage simulation pipelines~\citep{hartman2025controllable}, our method directly uses off-the-shelf MLIPs, making it easy to apply.

Our contributions are summarized as follows:
\begin{itemize}[leftmargin=*] \item \textbf{All-atom MOF generative framework:} We present an all-atom model for MOF structure prediction. By relaxing rigid-body assumptions, \ours{} yields a $85.61\%$ relative increase in match rate and a $34.32\%$ relative reduction in RMSD on the BW dataset~\citep{boyd2019data} compared to the best prior method. We also observe scaling behavior suggesting further gains with scale.
\item \textbf{Feynman-Kac steering with MLIPs:} We introduce MLIP-guided Feynman-Kac steering to improve structural quality and stability in unconstrained all-atom generation. It improves MOF validity by $10.9$--$28.7\%$, structural validity by $12.2$--$17.3\%$, and reduces formation energy error by $62.7$--$86.5\%$ across MLIP choices. 
\item \textbf{MOF-adsorbate structure prediction:} \ours{} is the first model to jointly predict MOF-adsorbate structures. On ODAC25~\citep{sriram2025open}, it generates diverse adsorption configurations with much lower wall-clock time than GCMC. When combined with MLIP relaxation, it can also find configurations with lower adsorption energies than the reference structures, which is important for accurate adsorption energy estimation~\citep{lan2023adsorbml}. \end{itemize}
\section{Related work} \label{related_work}

\subsection{Machine learning for crystal structure prediction} 
Most work on crystal structure prediction has focused on inorganic crystals. DiffCSP~\citep{jiao2024crystal} pioneered this direction by jointly denoising lattice and coordinate information with a periodic E(3)-equivariant architecture. Subsequent works improved symmetry handling: EquiCSP~\citep{linequivariant} accounts for lattice permutation symmetry, while DiffCSP++~\citep{jiaospace} and SymmCD~\citep{levysymmcd} incorporate space-group constraints. Beyond diffusion, other generative frameworks have also been explored: FlowMM~\citep{millerflowmm} uses equivariant flow matching, CrysBFN~\citep{wu2025periodic} applies a periodic E(3) Bayesian flow network, and OMatG~\citep{hollmeropen} adapts the stochastic interpolant framework to improve performance further.

In contrast, structure prediction for MOFs is more challenging because of their modular nature and larger system sizes. Existing methods address this complexity with a rigid-body assumption: MOFFlow~\citep{kim2025mofflow} predicts building-block rotations and translations via Riemannian flow matching, MOF-BFN~\citep{jiao2025mof} operates on fractional coordinates to enforce periodicity, and MOF-LLM~\citep{pan2026enhancing} uses progressive training with a large language model. Although MOFFlow-2~\citep{kim2025flexible} relaxes this assumption by incorporating torsion angles around rotatable bonds, it still keeps local geometries (bond lengths and angles) generated by cheminformatics tools fixed. In this work, we introduce an all-atom model that relaxes these constraints to enable fully flexible structure prediction.

\subsection{Generative models for molecular complexes} 
Recent generative models have increasingly addressed interactions within molecular complexes. In protein structure prediction, AlphaFold 1 and 2~\citep{senior2020improved, jumper2021highly} focused on isolated proteins, whereas AlphaFold 3~\citep{Abramson2024} extended prediction to complexes involving proteins, nucleic acids, and small molecules; subsequent models such as Boltz~\citep{wohlwend2025boltz, passaro2025boltz}, Chai-1~\citep{chai2024chai}, Protenix~\citep{bytedance2025protenix}, and SeedFold~\citep{yi2025seedfold} further advanced complex modeling. Likewise, in heterogeneous catalysis, generative models explicitly capture catalyst-adsorbate interactions~\citep{kolluru2024adsorbdiff, kim2026catflow}. Our approach extends MOF structure prediction in a similar spirit by jointly modeling the full MOF-adsorbate system.
\section{Method} \label{method}
\subsection{Problem formulation}
In this work, we study structure prediction for MOFs. Specifically, we predict the atomic coordinates of an MOF-adsorbate system given the underlying molecular graphs of the building blocks (organic linkers and metal ions) and, optionally, the adsorbates (\Cref{fig:concept}). 

\textbf{Crystal representation.} A MOF structure $\mathcal{S}$ is defined by a unit cell on a three-dimensional lattice. We represent a unit cell with $N$ atoms as $\mathcal{S} = (\mathbf{A}, \mathbf{X}, \bm{\ell})$, where $\mathbf{A} = \{a_i\}_{i=1}^N \in \mathcal{A}^N$ denotes the atom types, with $\mathcal{A}$ the set of chemical elements, and $\mathbf{X} = \{x_i\}_{i=1}^N \in \mathbb{R}^{N \times 3}$ denotes the atomic coordinates. The lattice is given by a matrix $L = (l_1, l_2, l_3) \in \mathbb{R}^{3 \times 3}$ whose columns are the lattice vectors. We parameterize it by $\bm{\ell} = (\bm{d}, \bm{\phi})$, where $\bm{d} = (a,b,c) \in \mathbb{R}_+^3$ are the edge lengths and $\bm{\phi} = (\alpha,\beta,\gamma) \in [0^\circ, 180^\circ]^3$ are the inter-edge angles. The corresponding infinite periodic crystal is $\{(a_n, x_n + Lm) \,\big|\, n \in [1, N],\, m \in \mathbb{Z}^3\}$, where $m = (m_1, m_2, m_3)^\top$ indexes lattice translations.

\textbf{Building block representation.} We represent the MOF-adsorbate system as a set of molecular graphs, denoted collectively by $\mathcal{B}=\{\mathcal{G}\}_{i=1}^M$, where $\mathcal{G}_i$ denotes the $i$-th molecular graph with atoms as vertices and bonds as edges. This set includes the MOF building blocks and, when present, the adsorbates; for bare MOFs, the adsorbate set is empty.

\subsection{Variational flow matching for MOFs} \label{method:vrm}

In this work, we adopt variational flow matching~\citep[VFM;][]{eijkelboom2024variational}, which enables training with an $L^1$ objective and empirically improves convergence. VFM is a generative modeling framework that learns to transform samples from a prior distribution $p_0(\mathbf{z}_0)$ to a target data distribution $q$ by predicting clean data from noisy observations, conditioned on the building blocks $\mathcal{B}$. 

\textbf{Conditional probability path.} We apply VFM to jointly model the atomic coordinates $\mathbf{X}$ and lattice parameters $\bm{\ell}$; for notational simplicity, let $\mathbf{z} = (\mathbf{X}, \bm{\ell}) \in \mathbb{R}^{N \times 3} \times \mathbb{R}^{6}$. We define a conditional probability path $p_t(\mathbf{z} \mid \mathbf{z}_1)$ that linearly interpolates between a prior sample $\mathbf{z}_0 \sim p_0$ at $t=0$ and a data sample $\mathbf{z}_1 \sim q$ at $t=1$, namely $\mathbf{z}_t = (1-t)\mathbf{z}_0 + t\mathbf{z}_1$.\footnote{We omit conditioning on building blocks $\mathcal{B}$ for brevity.}

\textbf{Training and inference.} Given $\mathbf{z}_t$, VFM learns a variational posterior $q_t^{\theta}(\mathbf{z}_1 \mid \mathbf{z}_t)$ that approximates the true posterior $p_{1|t}(\mathbf{z}_1 \mid \mathbf{z}_t)$. Training minimizes the expected negative log-likelihood:
\begin{equation}
    \label{eq:vfm}
    \mathcal{L}_{\text{VFM}}(\theta)
    = - \mathbb{E}_{t, \mathbf{z}_t, \mathbf{z}_1} \bigl[
    \log q_t^{\theta}(\mathbf{z}_1 \mid \mathbf{z}_t) \bigr],
\end{equation}
where $t \sim \mathcal{U}[0,1]$, $\mathbf{z}_1 \sim q$, and $\mathbf{z}_t \sim p_t(\mathbf{z} \mid \mathbf{z}_1)$. At inference time, we generate samples by solving the ordinary differential equation (ODE) $\mathrm{d}\bm{z}_t = \mathbf{v}_t(\mathbf{z}_t)\,\mathrm{d}t$, where the velocity field is given by $\mathbf{v}_t(\mathbf{z}_t) = (\bm{\mu}_t^{\theta}(\mathbf{z}_t) - \mathbf{z}_t)/(1-t)$.

\textbf{Parameterization.} We model the variational posterior $q_t^\theta(\mathbf{X}_1, \bm{\ell}_1 \mid \mathbf{X}_t, \bm{\ell}_t)$ as a Laplace distribution whose mean $\bm{\mu}_t^{\theta}$ is parameterized by the neural network as
    $\hat{\mathbf{X}}_1, \hat{\bm{\ell}}_1 = \bm{\mu}_t^{\theta}(\mathbf{X}_t, \bm{\ell}_t \mid \mathcal{B})$,
where $\hat{\mathbf{X}}_1$ and $\hat{\bm{\ell}}_1$ are the predicted clean Cartesian coordinates and lattice parameters, respectively. Under this parameterization, \Cref{eq:vfm} reduces to the $L^1$ reconstruction loss \citep{zaghen2025riemannian}:
\begin{equation}
\label{eq:loss}
\begin{aligned}
    \mathcal{L}(\theta) = \mathbb{E}_{\mathcal{S}_1, \mathcal{S}_t, t}
        & \Big[ \lambda_{\text{coord}} \| \hat{\mathbf{X}}_1 - \mathbf{X}_1 \|_1
        + \lambda_{\text{lattice}} \| \hat{\bm{\ell}}_1 - \bm{\ell}_1 \|_1 \Big],
\end{aligned}
\end{equation}
where $\mathcal{S}_1 = (\mathbf{A}, \mathbf{X}_1, \bm{\ell}_1) \sim q(\mathcal{S})$, $\mathcal{S}_t = (\mathbf{A}, \mathbf{X}_t, \bm{\ell}_t)$, $t \sim \mathcal{U}[0,1]$, and $\lambda_{\text{coord}}$ and $\lambda_{\text{lattice}}$ are weights. 

\textbf{Prior distribution.}
We define prior distributions over the atomic coordinates $\mathbf{X}$ and lattice parameters $\bm{\ell} = (\bm{d}, \bm{\phi})$. For the coordinates, we use a centered Gaussian prior $\mathbf{X}_0 = \mathbf{P}\bm{\varepsilon}$, where $\bm{\varepsilon} \sim \mathcal{N}(0, \mathbf{I})$ and $\mathbf{P} = \mathbf{I} - \frac{1}{N}\mathbf{1}\mathbf{1}^{\top}$ enforces a zero center of mass. For the lattice parameters, the lengths $\bm{d}$ are sampled independently from log-normal distributions, $\bm{d} \sim \operatorname{LogNormal}(\bm{\mu}, \bm{\sigma})$, where $\bm{\mu}$ and $\bm{\sigma}$ are estimated by maximum likelihood. We sample lattice angles $\bm{\phi}$ uniformly from $[60^\circ, 120^\circ]$~\citep{millerflowmm}.

\subsection{Model architecture} \label{method:model}

The denoising network, $\bm{\mu}_t^{\theta}(\bm{X}_t, \bm{\ell}_t \vert \mathcal{B})$, retains a scalable DiT backbone~\citep{peebles2023scalable, joshi2025allatom} while augmenting attention with a pairwise representation that encodes local distances and connectivity within MOF building blocks, inspired by AlphaFold 3~\citep{Abramson2024}. The network comprises three stages: an \textit{encoder} that constructs features from the building blocks and noisy inputs, a deep \textit{trunk} that models interactions, and a \textit{decoder} that predicts $\bm{X}_1$ and $\bm{\ell}_1$. See \Cref{fig:model_architecture} for an overview and \Cref{app:architecture} for pseudocode.

\textbf{Encoder.}
The encoder initializes single-atom ($\bm{Q} \in \mathbb{R}^{N \times D_Q}$) and pairwise ($\bm{P} \in \mathbb{R}^{N \times N \times D_P}$) representations from the building blocks $\mathcal{B}$ and the noisy inputs $\bm{X}_t$ and $\bm{\ell}_t$. First, we construct initial single features $\bm{C}$ from atom types, formal charges, and local reference coordinates $\bm{X}_{\text{local}}$ derived from RDKit~\cite{rdkit}. In parallel, we initialize $\bm{P}$ by featurizing the internal geometry of the building blocks. Specifically, we compute local displacement vectors $\bm{D}_{ij} = \bm{X}_{\text{local},i} - \bm{X}_{\text{local},j}$ and Euclidean distances $\|\bm{D}_{ij}\|$, and embed them together with bond types and block-adjacency masks indicating whether two atoms belong to the same building block. We then condition $\bm{P}$ on $\bm{C}$ using an MLP, while forming $\bm{Q}$ by adding projections of the noisy global coordinates $\bm{X}_t$ and lattice parameters $\bm{\ell}_t$ to $\bm{C}$. Finally, these features are processed by a shallow $\text{DiT}_{\text{encoder}}$, where $\bm{P}$ acts as an additive bias on the attention logits (see \Cref{alg:attention} in \Cref{app:architecture}).

\textbf{Trunk.}
The trunk performs the main interaction modeling through a deep stack of DiT blocks. We first project the encoder outputs into higher-dimensional representations $\bm{S}$ and $\bm{Z}$ using an MLP. Before the transformer pass, we update the pairwise bias $\bm{Z}$ with single-atom context via
$
\bm{Z}_{ij} \leftarrow \bm{Z}_{ij} + \operatorname{Linear}(\bm{S}_i) + \operatorname{Linear}(\bm{S}_j).
$
The updated $\bm{Z}$ then serves as the structural attention bias throughout the trunk. After the trunk computation, we apply a global residual update to the encoder features: $\bm{Q} \leftarrow \bm{Q} + \operatorname{Linear}(\bm{S}')$.

\textbf{Decoder.}
The decoder maps the updated features $\bm{Q}$ to the denoised state $\hat{\bm{X}}_1$ and $\hat{\bm{\ell}}_1$. After a $\text{DiT}_{\text{decoder}}$, we use two parallel heads, each consisting of a LayerNorm and a linear layer, to generate the outputs. The coordinate head projects per-atom features to $\hat{\bm{X}}_1 \in \mathbb{R}^{N \times 3}$, while the lattice head applies global average pooling followed by a projection to $\hat{\bm{\ell}}_1 \in \mathbb{R}^{6}$.

\subsection{Feynman-Kac steering with MLIP} \label{method:fk_steering}

While all-atom generative models offer high flexibility, they can also produce structural distortions and incorrect bond lengths~\citep{jiang2025posex}. To mitigate this issue, we introduce Feynman-Kac steering guided by a machine learning interatomic potential (MLIP), which improves the structural fidelity and stability of generated molecules at inference time. The advantage of this approach is simplicity: off-the-shelf MLIPs greatly improve structural quality without hand-designed potential functions~\citep{wohlwend2025boltz} or expensive domain-specific simulations~\citep{hartman2025controllable}.

\textbf{MLIP-guided Feynman-Kac steering.} Intuitively, we bias toward structures that are energetically favorable under the MLIP. Concretely, we sample from an exponentially tilted distribution:
\begin{equation}
    p_{\text{target}}(\bm{X}, \bm{\ell} \mid \mathcal{B}) 
    \propto p_{\theta}(\bm{X}, \bm{\ell} \mid \mathcal{B}) \exp\left(-\lambda E(\mathcal{S})\right),
\end{equation}
where $E(\cdot)$ is the energy estimated by the MLIP and $\lambda$ controls the steering strength. In Feynman-Kac steering, a \emph{potential} is an intermediate scoring function used to reweight and resample particles at noisy states along the trajectory; larger values indicate particles that are more likely to yield desirable final samples~\citep{singhalgeneral, hartman2025controllable}. Different choices of potential correspond to different ways of converting the MLIP signal into intermediate resampling scores. In our experiments, we use the immediate potential, which depends only on the current timestep, unless noted otherwise. Definitions of the different potentials and the full algorithm are provided in \Cref{app:fk_algorithm}.

\textbf{SDE reformulation.} To enable particle resampling for Feynman-Kac steering, we convert the ODE induced by the learned velocity field $\mathbf{v}(\bm{X}_t, t)$ into the SDE
\begin{equation} \label{eq:equivalent_sde}
\mathrm{d}\bm{X}_t
=
\Big(
\mathbf{v}(\bm{X}_t, t) + \tfrac{1}{2} g(t)^2\, s(\bm{X}_t, t)
\Big)\mathrm{d}t
+ g(t)\, \mathrm{d}\bm{W}_t,
\end{equation}
where $\bm{W}_t$ is a standard Wiener process, $g(t)$ is the diffusion coefficient~\citep{ma2024sit}, and $s(\bm{X}_t, t)$ is the score function. For a centered Gaussian prior $\bm{X}_0 \sim \mathcal{N}_C(\bm{0}, \bm{I})$, the score is given directly by $\mathbf{v}$:
$s(\bm{X}_t, t) = (t\,\mathbf{v}(\bm{X}_t, t) - \bm{X}_t)/(1 - t)$.
\section{MOF-only structure prediction} \label{exp:mof_sp}

\subsection{Experimental setup} \label{exp:setup}

\textbf{Data preprocessing.} We apply a unified preprocessing pipeline to both datasets. We first use MOFid~\citep{bucior2019identification} to decompose each MOF into nodes, linkers, solvents, and adsorbates. We then validate the organic linkers with RDKit and adjust atomic coordinates to ensure each building block is spatially contiguous. Details for constructing node templates are provided in \Cref{app:train_details}.

\textbf{Training and hyperparameters.} We train all variants with a three-stage curriculum that progressively increases system size. The first two stages restrict the training set by maximum atom count, and the final stage fine-tunes on the full dataset. We train three model sizes: \ours{}-S (38.4M parameters), \ours{}-M (129M), and \ours{}-L (491M). Full training details are provided in \Cref{app:train_details}, and architecture, curriculum, and deterministic prediction comparisons in \Cref{app:ablation_details}.

\subsection{Match rate \& RMSD} \label{exp:mr_rmsd}

\begin{wrapfigure}{r}{0.42\textwidth}
    \centering
    \vspace{-10pt}
    \resizebox{0.42\textwidth}{!}{%
    \begin{tabular}{ccc}
        \toprule
        MOFFlow-2
        & \ours{}
        & Reference \\
        \midrule
        \includegraphics[width=0.3\linewidth]{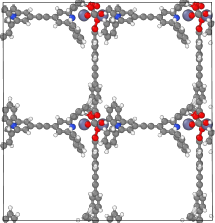} &
        \includegraphics[width=0.3\linewidth]{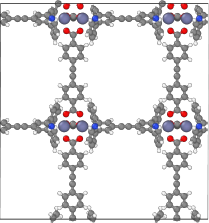} & 
        \includegraphics[width=0.3\linewidth]{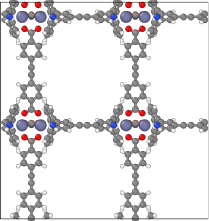} \\
        
        \arrayrulecolor{black!40}\midrule
        \arrayrulecolor{black}
        \includegraphics[width=0.3\linewidth]{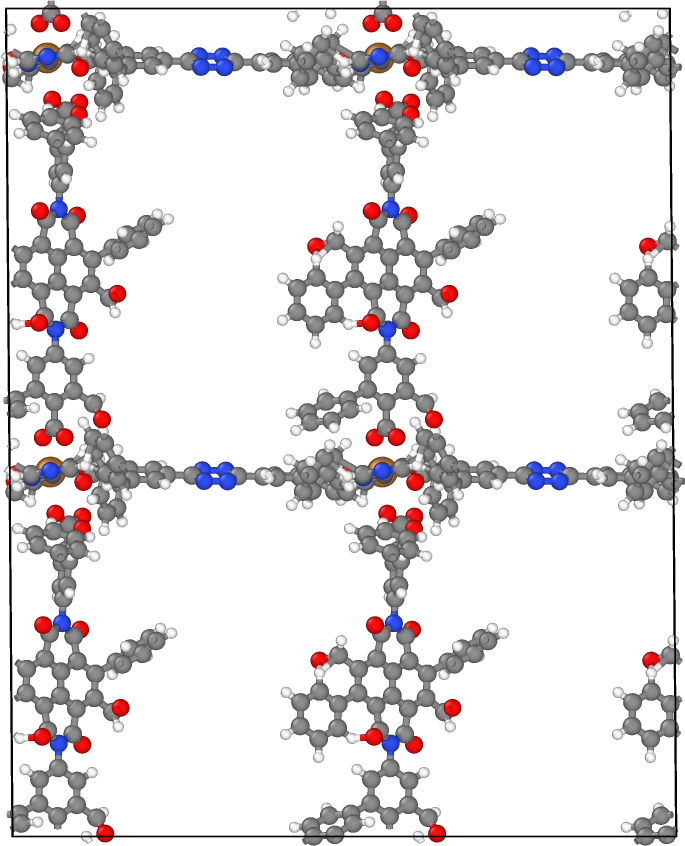} &
        \includegraphics[width=0.3\linewidth]{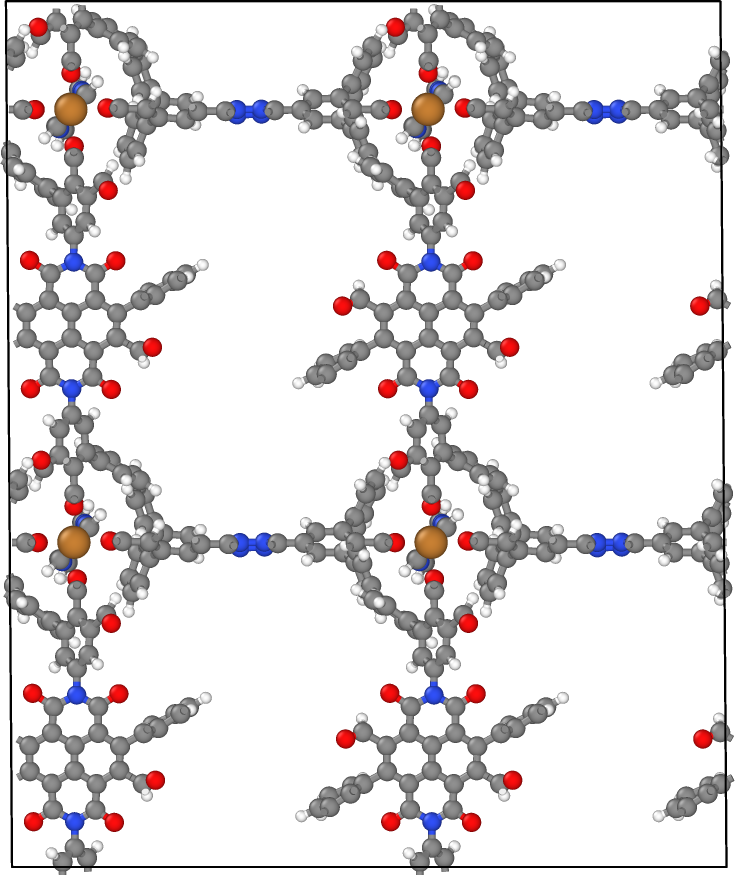} &
        \includegraphics[width=0.3\linewidth]{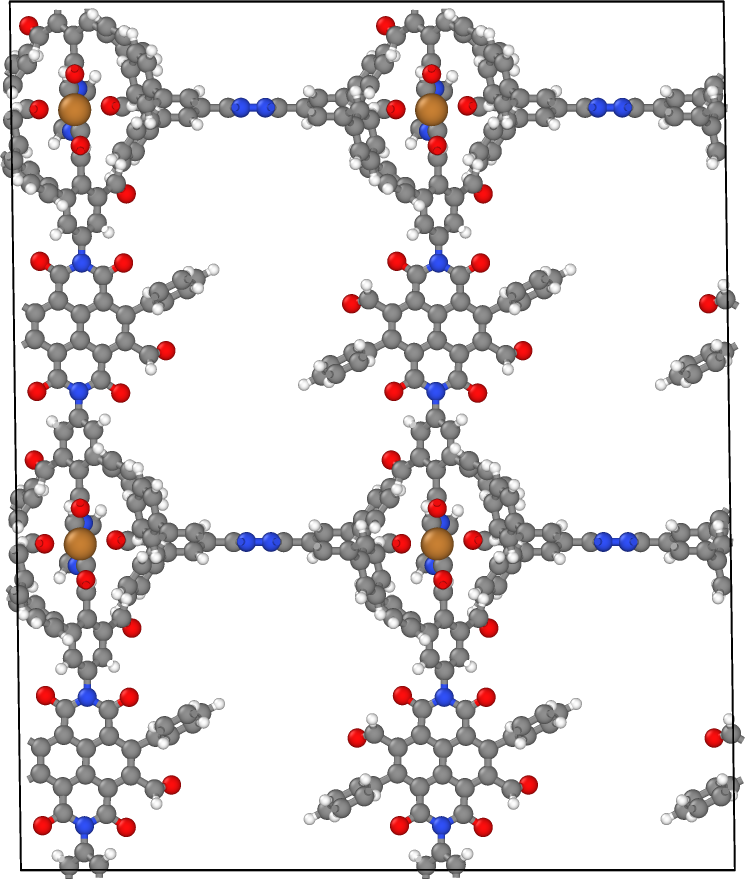} \\
        \bottomrule
    \end{tabular}%
    }
    \caption{\textbf{Visualizations.} Structures in $2\times2\times2$ supercells. \ours{}-M better matches the ground truth and reduces atomic clashes near metal nodes.}
    \vspace{-20pt}
    \label{fig:bwdb_visualization}
\end{wrapfigure}

\textbf{Overview.} We evaluate our model on the BW dataset for MOF structure prediction from the 2D molecular graphs of building blocks $\mathcal{B}$. We use the 8:1:1 split of \citet{kim2025flexible}. At inference time, we generate samples with an ODE solver using 10 integration steps and report results for 1 and 5 samples, evaluating the best-matching candidate for both cases.

\textbf{Metrics.} We assess performance using match rate (MR) and root mean squared deviation (RMSD), computed with \texttt{pymatgen}'s \texttt{StructureMatcher}. The matcher compares structures based on specified tolerances for site positions ($\operatorname{stol}$), fractional lengths ($\operatorname{ltol}$), and angles ($\operatorname{angle\_tol}$), returning 1 if the structures match within these limits and 0 otherwise. RMSD is computed only for successfully matched pairs. We use $\operatorname{stol}$ values of 0.5 (a typical value for CSP tasks) and 0.3 (the stricter default), $\operatorname{ltol}=0.3$, and $\operatorname{angle\_tol}=10$.

\textbf{Baselines.} We compare our method against both optimization- and learning-based baselines. The optimization-based baselines include random search (RS) and evolutionary algorithms (EA), which iteratively propose candidates until an energetically favorable structure is found (implemented with \texttt{CrySPY}~\citep{yamashita2021cryspy}). We also include DiffCSP, a general CSP baseline for inorganic materials, and MOF-specific baselines, namely MOFFlow~\citep{kim2025mofflow}, MOFFlow-2~\citep{kim2025flexible}, and MOF-BFN~\citep{jiao2025mof}.

The original MOFFlow and MOF-BFN implementations assume access to ground-truth local coordinates, which is unrealistic in practice. For a fair comparison, we retrain both models following \citet{kim2025flexible}. During training, we use matched coordinates obtained by aligning nodes and RDKit linkers to the ground truth; at inference time, we use node templates and direct RDKit predictions.

\textbf{Results.} As shown in \Cref{table:bwdb}, \ours{} consistently outperforms all baselines across every metric. Although MOFFlow and MOF-BFN are smaller, they predict only rigid-body transformations of building blocks rather than all-atom coordinates, which limits flexibility and reconstruction accuracy. In the 1-sample case under the stricter setting ($\operatorname{stol}=0.3$), at a comparable parameter count, \ours{}-M (129M) yields a $74.76\%$ relative increase in match rate and a $30.94\%$ relative reduction in RMSD relative to MOFFlow-2 (146M). \ours{}-L performs even better, yielding a $85.61\%$ increase in match rate and a $34.32\%$ reduction in RMSD. This advantage is evident in \Cref{fig:bwdb_visualization}, where MOFFlow-2 exhibits steric clashes around metal nodes, whereas \ours{} faithfully reconstructs the ground-truth structures.

\begin{table}[t]
\centering\small
\caption{\textbf{Comparison of structure prediction accuracy.}
MR is the match rate, and RMSD is the root mean squared deviation.
$\operatorname{stol}$ is the site-tolerance, where smaller values impose stricter matching. Param. is the number of parameters.
\textbf{Bold} indicates the best result and \underline{underlined} indicates the second-best.
Random sampling and evolutionary algorithm baselines achieve $0\%$ match rate at all tolerances with 20 samples and are omitted.
\ours{} achieves best performance, with larger gains under stricter tolerance, highlighting the importance of all-atom modeling.
}
\label{table:bwdb}
\setlength{\tabcolsep}{\bwdbtabcolsep}
\renewcommand{\arraystretch}{\bwdbtablestretch}
\resizebox{\columnwidth}{!}{%
\begin{tabular}{
    l
    c
    S[table-format=2.2, detect-weight, mode=text]
    S[table-format=1.4, detect-weight, mode=text]
    S[table-format=2.2, detect-weight, mode=text]
    S[table-format=1.4, detect-weight, mode=text]
    S[table-format=2.2, detect-weight, mode=text]
    S[table-format=1.4, detect-weight, mode=text]
    S[table-format=2.2, detect-weight, mode=text]
    S[table-format=1.4, detect-weight, mode=text]
}
\toprule
 & & \multicolumn{4}{c}{1 Sample} & \multicolumn{4}{c}{5 Samples} \\
\cmidrule(lr){3-6} \cmidrule(lr){7-10}
 & & \multicolumn{2}{c}{$\operatorname{stol}=0.3$} & \multicolumn{2}{c}{$\operatorname{stol}=0.5$}
 & \multicolumn{2}{c}{$\operatorname{stol}=0.3$} & \multicolumn{2}{c}{$\operatorname{stol}=0.5$} \\
\cmidrule(lr){3-4} \cmidrule(lr){5-6} \cmidrule(lr){7-8} \cmidrule(lr){9-10}
\textbf{Model} & {Param.}
 & {MR (\%) $\uparrow$} & {RMSD $\downarrow$} & {MR (\%) $\uparrow$} & {RMSD $\downarrow$}
 & {MR (\%) $\uparrow$} & {RMSD $\downarrow$} & {MR (\%) $\uparrow$} & {RMSD $\downarrow$} \\
\midrule
DiffCSP   & {12.3M} & 0.01 & 0.1554 & 0.23 & 0.3896 & 0.08 & 0.1299 & 0.87 & 0.3982 \\
MOFFlow   & {22.5M} & 5.28 & 0.2036 & 21.93 & 0.3329 & 8.68 & 0.2039 & 32.71 & 0.3290 \\
MOF-BFN   & {11.2M} & 6.96 & 0.1979 & 27.66 & 0.3234 & 12.11 & 0.1950 & 40.15 & 0.3131 \\
MOFFlow-2 & {146M}  & 8.20 & 0.1894 & 28.71 & 0.3094 & 15.98 & 0.1842 & 43.95 & 0.2925 \\
\midrule
\rowcolor{ourscolor}
\ours{}-M & {129M} & \underline{14.33} & \underline{0.1308} & \underline{37.88} & \underline{0.2542}
                  & \underline{26.67} & \underline{0.1218} & \underline{59.77} & \underline{0.2329} \\
\rowcolor{ourscolor}
\ours{}-L & {491M} & \textbf{15.22} & \textbf{0.1244} & \textbf{39.82} & \textbf{0.2498}
                  & \textbf{28.97} & \textbf{0.1168} & \textbf{62.04} & \textbf{0.2243} \\
\bottomrule
\end{tabular}%
}
\vspace{\bwdbtablebottomskip}
\end{table}

\subsection{Property evaluation} \label{exp:property}

\begin{wraptable}{r}{\propertytablewidth}
    \centering\small
    \vspace{-10pt}
    \caption{\textbf{Property evaluation.} RMSE between ground-truth and generated MOFs for key properties computed via \texttt{Zeo++}. \ours{} shows lower errors than MOFFlow-2, highlighting the benefit of all-atom modeling.}
    \label{tab:property_rmsd}
    \resizebox{\linewidth}{!}{%
        \begin{tabular}{lrrrr}
        \toprule
         & VSA $\downarrow$ & GSA $\downarrow$ & AV $\downarrow$ & UCV $\downarrow$ \\
        \midrule
        MOFFlow-2 & 312.8 & 443.6 & 485.0 & 427.1 \\
        \rowcolor{ourscolor} \ours{}-M & \underline{216.1} & \underline{252.0} & \textbf{176.4} & \textbf{148.8} \\
        \rowcolor{ourscolor} \ours{}-L & \textbf{202.0} & \textbf{232.9} & \underline{193.3} & \underline{167.3} \\
        \midrule
         & VF $\downarrow$ & PLD $\downarrow$ & LCD $\downarrow$ & DST $\downarrow$ \\
        \midrule
        MOFFlow-2 & 0.0445 & 1.3751 & 1.5361 & 0.0312 \\
        \rowcolor{ourscolor} \ours{}-M & \underline{0.0214} & \underline{0.9025} & \underline{0.9594} & \underline{0.0170} \\
        \rowcolor{ourscolor} \ours{}-L & \textbf{0.0197} & \textbf{0.8715} & \textbf{0.8936} & \textbf{0.0169} \\
        \bottomrule
        \end{tabular}%
    }
    \vspace{-5pt}
\end{wraptable}

\textbf{Overview.} We evaluate whether our model preserves key structural properties relevant to downstream applications such as gas storage and catalysis. For each generated structure, we measure volumetric surface area (VSA; $\mathrm{m^2/cm^3}$), gravimetric surface area (GSA; $\mathrm{m^2/g}$), largest cavity diameter (LCD; \AA), pore limiting diameter (PLD; \AA), accessible volume (AV; \AA$^3$), unit-cell volume (UCV; \AA$^3$), density (DST; $\mathrm{g/cm^3}$), and void fraction (VF).

\textbf{Metrics \& baselines.} We compare against MOFFlow-2 using its default sampling configuration of 50 ODE steps, while \ours{} uses 10 integration steps. All properties are computed with \texttt{Zeo++}~\citep{willems2012algorithms}. We report the RMSE between predicted and ground-truth properties.

\textbf{Results.} As shown in \Cref{tab:property_rmsd}, both \ours{} variants achieve lower property errors than MOFFlow-2 across all reported metrics. The largest gains are in accessible volume (AV) and unit-cell volume (UCV): relative to MOFFlow-2, \ours{}-M reduces RMSE from 485.0 to 176.4 for AV and from 427.1 to 148.8 for UCV. \ours{}-L performs best on the other six properties, namely VSA, GSA, VF, PLD, LCD, and density. Overall, these results suggest that all-atom modeling preserves structural properties better than torsion-based parameterizations with fixed bond constraints, and that scaling further improves most property estimates.

\subsection{Scaling \ours{}} \label{exp:scaling_law}

\definecolor{ourspink}{HTML}{D81B60}
\definecolor{flowblue}{HTML}{1E88E5}
\newcommand{\hlc}[3]{\raisebox{0pt}[0pt][0pt]{\colorbox{#1!#2}{#3}}}

\begin{figure}[t]
    \centering
    \begin{minipage}[t]{0.48\linewidth}
        \centering
        \begin{subfigure}[b]{0.48\linewidth}
            \centering
            \includegraphics[width=\linewidth]{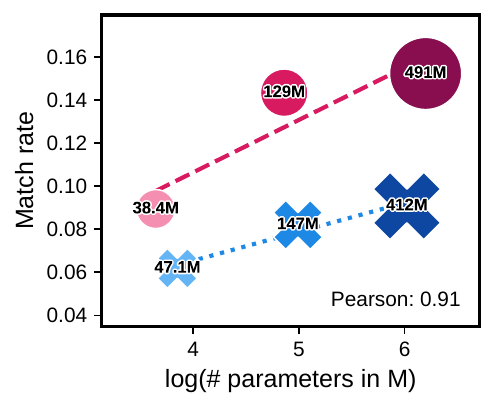}
            \vspace{-15pt}
            \label{fig:scaling_mr}
        \end{subfigure}%
        \hfill%
        \begin{subfigure}[b]{0.48\linewidth}
            \centering
            \includegraphics[width=\linewidth]{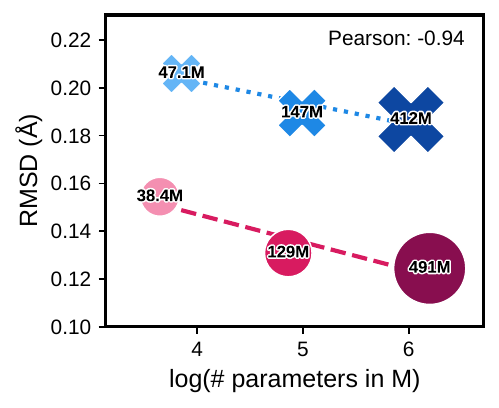}
            \vspace{-15pt}
            \label{fig:scaling_rmsd}
        \end{subfigure}
        \caption{\textbf{Scaling behavior of \ours{}.} Match rate (\textit{left}) and RMSD (\textit{right}) versus model size. \hlc{ourspink}{20}{\ours{}} achieves a higher match rate and lower RMSD across all model sizes than \hlc{flowblue}{20}{MOFFlow-2}. It also exhibits better scaling behavior, with Pearson correlations near $\pm 0.9$, indicating predictable gains with scale.}
        \label{fig:scaling_law}
    \end{minipage}%
    \hfill%
    \begin{minipage}[t]{0.48\linewidth}
        \centering
        \begin{subfigure}[b]{0.48\linewidth}
            \centering
            \includegraphics[width=\linewidth]{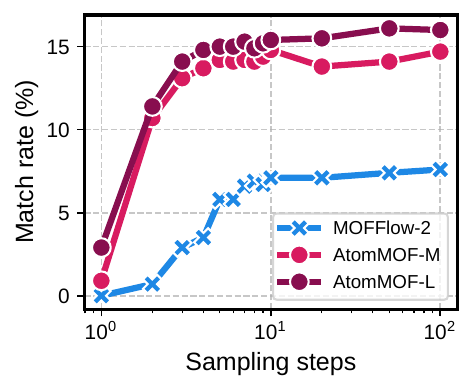}
            \vspace{-15pt}
            \label{fig:sampling_steps}
        \end{subfigure}%
        \hfill%
        \begin{subfigure}[b]{0.48\linewidth}
            \centering
            \includegraphics[width=\linewidth]{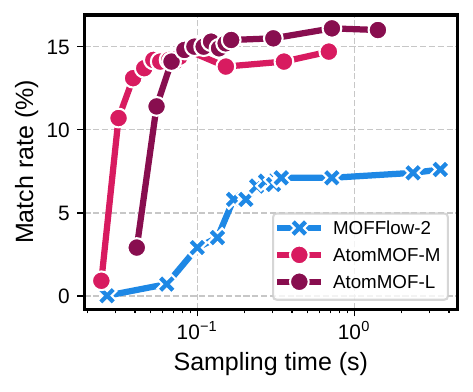}
            \vspace{-15pt}
            \label{fig:sampling_time}
        \end{subfigure}
        \caption{\textbf{Inference efficiency under scaling.} Match rate (\%) versus (\textit{left}) ODE sampling steps and (\textit{right}) wall-clock time. \ours{}-L outperforms \ours{}-M at the same step or time budget, and both surpass MOFFlow-2. \ours{} exceeds MOFFlow-2's peak match rate within a few steps and saturates by $10$ steps.}
        \label{fig:inference_efficiency}
    \end{minipage}
    \vspace{-1em}
\end{figure}

\begin{wrapfigure}{r}{\dimexpr\scalingfigwidth+\scalingfigsep\relax}
    \vspace{-10pt}
    \hfill
    \begin{minipage}{\scalingfigwidth}
    \centering
    \small
    \resizebox{\linewidth}{!}{%
    \begin{tabular}{cccc}
        \toprule
        Small
        & Medium
        & Large
        & Reference 
        \\
        \midrule
        \includegraphics[width=0.18\linewidth]{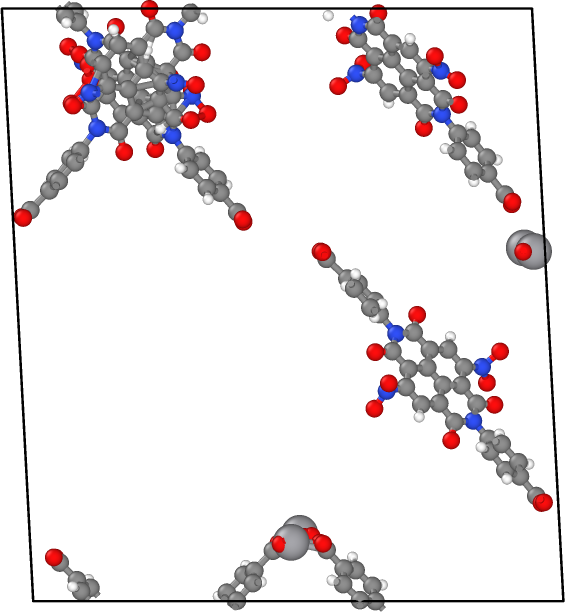} & 
        \includegraphics[width=0.18\linewidth]{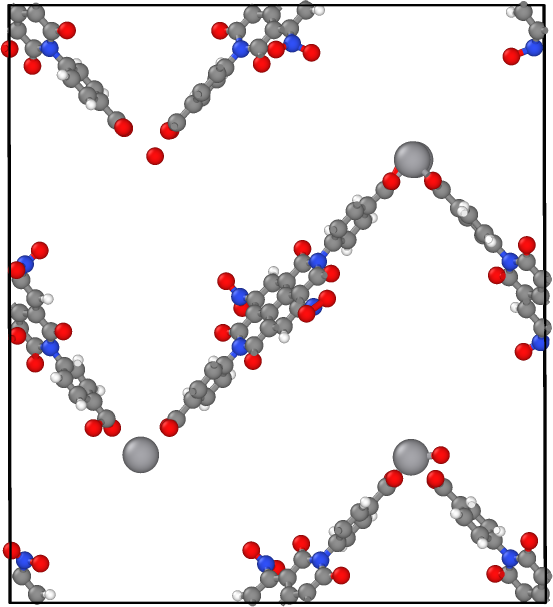} & 
        \includegraphics[width=0.18\linewidth]{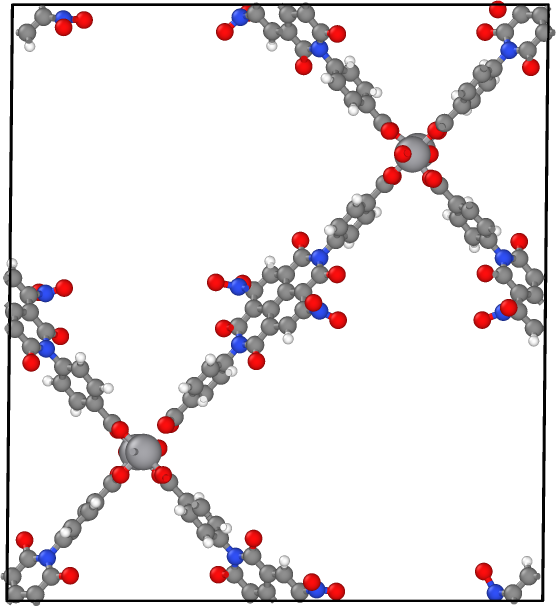} & 
        \includegraphics[width=0.18\linewidth]{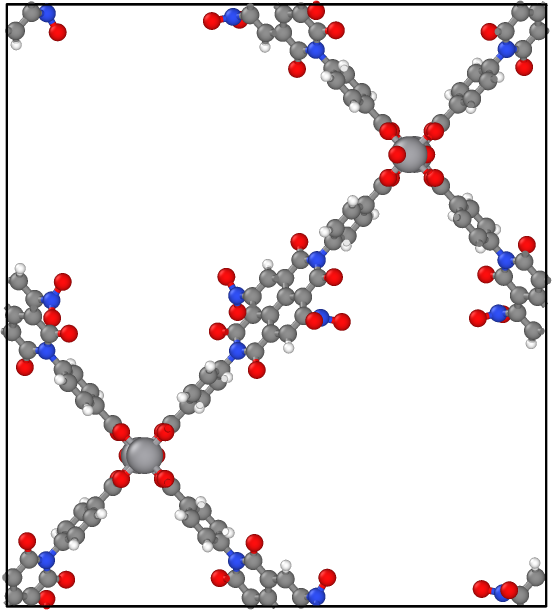} \\
        \bottomrule
    \end{tabular}%
    }
    \caption{\textbf{Effect of model scaling on predicted structures.} From left to right, larger models produce higher-quality structures that more closely match the reference, better recovering the reference topology and reducing steric clashes.}
    \label{fig:scaling_vis}
    \end{minipage}
    \vspace{-1.5em}
\end{wrapfigure}

\textbf{Parameter scaling.} \Cref{fig:scaling_law} shows consistent performance gains as we scale the model from 38.4M to 491M parameters. For comparison, we also train two MOFFlow-2 variants, a small model with 47.1M parameters and a large model with 412M parameters, to assess scaling relative to \ours{}. Both match rate and RMSD follow an approximately log-linear trend (Pearson $+0.91/-0.94$) with better scaling behavior than MOFFlow-2. All results use $\operatorname{stol}=0.3$, with 10 integration steps for \ours{} and 50 for MOFFlow-2. Qualitative improvements with scale are shown in \Cref{fig:scaling_vis}.

\textbf{Runtime scaling.} \Cref{fig:inference_efficiency} shows how match rate ($\operatorname{stol}=0.3$) varies with the number of sampling steps. We evaluate 1{,}000 randomly sampled test structures using 1 to 100 ODE integration steps on a single A100 80GB GPU. In \Cref{fig:inference_efficiency} (left), \ours{}-L outperforms \ours{}-M at every step count, indicating higher sample efficiency per integration step. In terms of wall-clock time (\Cref{fig:inference_efficiency}, right), \ours{}-M performs better at very small time budgets, but \ours{}-L becomes superior once runtime exceeds $0.0825$\,s. Both variants are far more efficient than MOFFlow-2: \ours{} exceeds its peak match rate ($\sim$7.6\%) in just 3 steps and largely saturates by 10 steps, reaching $15.4\%$ for \ours{}-L and $14.8\%$ for \ours{}-M, which supports our default of 10 integration steps. 

\subsection{Feynman-Kac steering with MLIP} \label{exp:feynman_kac}

\definecolor{myred}{HTML}{B85450}
\definecolor{mygreen}{HTML}{4A8B6F}
\newcommand{\yes}{\textcolor{mygreen}{\faCheck}}
\newcommand{\no}{\textcolor{myred}{\faTimes}}

\edef\fktablebegin{\noexpand\begin{wraptable}\ifx\fktablelines\empty\else[\fktablelines]\fi{r}{\noexpand\fktablewidth}}
\fktablebegin
    \centering
    \vspace{\fktabletopskip}
    \caption{\textbf{Effect of Feynman-Kac steering with different MLIPs.} Steering improves \texttt{MOFChecker} validity (\texttt{MOF}, \%) and structural validity (Struct., \%) and reduces formation energy error $\Delta E_f$ (eV/atom, by eSEN).}
    \label{tab:fk_steering_results}
    \setlength{\tabcolsep}{4pt}
    \resizebox{\fktablewidth}{!}{%
    \begin{tabular}{l c c c c}
        \toprule
        \textbf{Method} & FK & \texttt{MOF} $\uparrow$ & Struct. $\uparrow$ & $\Delta E_f$ $\downarrow$ \\
        \midrule
        No steering & \no  & 34.16 & 81.70 & 0.126 \\
        \arrayrulecolor{black!40}\midrule
        \arrayrulecolor{black}
        \rowcolor{ourscolor}
        eSEN~\citep{sriram2025open}     & \yes & 37.87 & 91.70 & 0.020 \\
        \rowcolor{ourscolor}
        UMA~\citep{wood2025family}      & \yes & 43.98 & 94.90 & 0.017 \\
        \rowcolor{ourscolor}
        Orb-v3~\citep{rhodes2025orb}      & \yes & 41.47 & 95.80 & 0.036 \\
        \rowcolor{ourscolor}
        SevenNet~\citep{kim2026optimizing} & \yes & 41.52 & 95.50 & 0.047 \\
        \bottomrule
    \end{tabular}%
    }
    \vspace{\fktablebottomskip}
\end{wraptable}

\textbf{FK steering improves stability and validity.}
We apply Feynman-Kac steering with four MLIPs as guidance potentials: eSEN~\citep{sriram2025open}, UMA~\citep{wood2025family}, Orb-v3~\citep{rhodes2025orb}, and SevenNet-Omni~\citep{kim2026optimizing}. To avoid bias, we use held-out eSEN to evaluate $\Delta E_f$. We set $\lambda=2.0$, resampling interval 5, start time $t=0.8$, and the immediate potential from \Cref{tab:potentials}, using 16 particles for eSEN and UMA and 8 for Orb and SevenNet due to compute limits. As shown in \Cref{tab:fk_steering_results}, steering improves all metrics across guidance MLIPs: $\Delta E_f$ decreases by $62.7$--$86.5\%$, \texttt{MOFChecker} validity~\citep{jablonka2023mofchecker} increases by $10.9$--$28.7\%$, and structural validity by $12.2$--$17.3\%$, showing robustness to the MLIP choice. See \Cref{app:fk_details} for settings, metric definitions, compute overhead, and comparison with best-of-N, and \Cref{app:inorganic_csp_steering} for extension to inorganic crystals.

\textbf{Sensitivity analysis.} \Cref{fig:fk_hyperparam} evaluates steering sensitivity on 1,000 random test samples using UMA guidance and eSEN evaluation. Increasing the number of particles and steering strength ($\lambda$) consistently improves both stability and validity. The immediate potential mode performs best, consistent with \citet{hartman2025controllable}. Importantly, applying steering later in the denoising trajectory yields greater improvements, leveraging higher accuracy of MLIPs as structures denoise. Additional details are in \Cref{app:fk_details}.

\begin{figure}[t]
    \centering
    \iffkhyperparamwide
        \includegraphics[width=\fkhyperparamwidth]{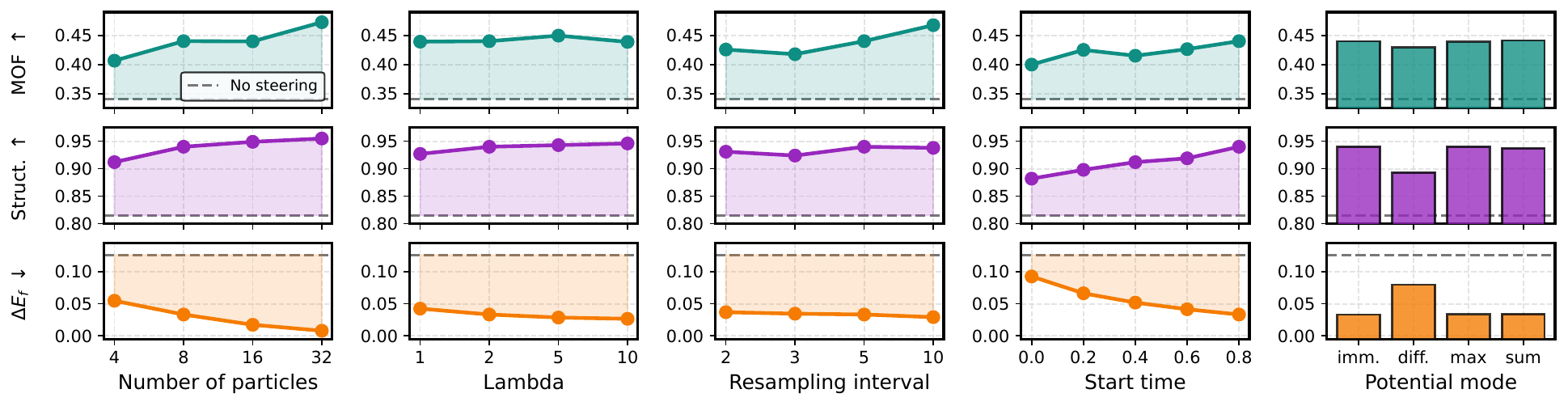}
    \else
        \includegraphics[width=\fkhyperparamwidth]{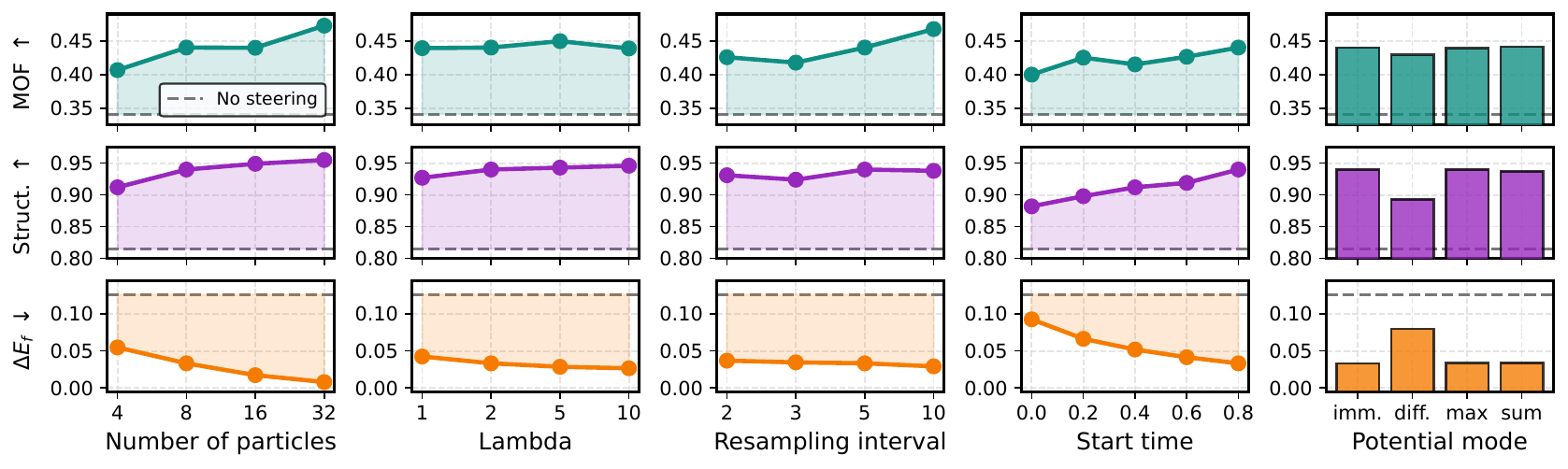}
    \fi
    \caption{\textbf{Sensitivity analysis of MLIP-guided Feynman-Kac steering.} Using UMA for steering, we ablate particle count, steering strength ($\lambda$), resampling interval, start time, and potential mode, and report \texttt{MOFChecker} validity (\%, \textit{top}), structural validity (\%, \textit{middle}), and eSEN formation energy error $\Delta E_f$ (eV/atom, \textit{bottom}). Increasing particle count and $\lambda$ generally improves validity and stability. Notably, later start times perform best, leveraging accurate MLIP predictions as the structures denoise.}
    \label{fig:fk_hyperparam}
    \vspace{-5pt}
\end{figure}

\section{MOF-adsorbate structure prediction} \label{exp:mof_ads_sp}

\begin{figure}[t]
    \centering
    \begin{minipage}[t]{0.44\linewidth}
        \centering
        \vspace{0pt} 
        \includegraphics[width=\linewidth,height=\odacplotheight]{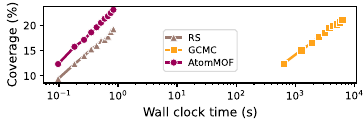}
        \caption{\textbf{Coverage vs.\ wall-clock time.} Coverage (\%) of reference adsorption configurations versus wall-clock time (log scale) for \ours{}, random sampling (RS), and GCMC. \ours{} attains the highest coverage at the lowest runtime.}
        \label{fig:coverage_vs_time}
    \end{minipage}%
    \hfill%
    \nextfloat
    \begin{minipage}[t]{0.50\linewidth}
        \centering
        \vspace{0pt} 
        \begin{subfigure}[b]{0.48\linewidth}
            \centering
            \includegraphics[width=\linewidth,height=\odacplotheight]{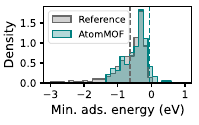}%
            \label{fig:1_sample_energy}
        \end{subfigure}%
        \hfill%
        \begin{subfigure}[b]{0.48\linewidth}
            \centering
            \includegraphics[width=\linewidth,height=\odacplotheight]{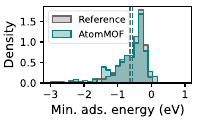}%
            \label{fig:10_sample_energy}
        \end{subfigure}
        \caption{\textbf{Adsorption energy histogram.} Histogram of minimum adsorption energies for each MOF-adsorbate system with 1 generated sample (\textit{left}) and 10 generated samples (\textit{right}). Vertical lines indicate the mean. The distribution closely matches the reference.}
        \label{fig:ads_energy_hist}
    \end{minipage}
    \vspace{-1em}
\end{figure}

We evaluate \ours{} on ODAC25 through two questions. First, under a fixed computational budget, how well does \ours{} recover dataset adsorption configurations (\Cref{exp:odac25_cov})? Second, combied with MLIP relaxation, can it identify lower-energy configurations than those in the reference dataset (\Cref{exp:odac25_ads})? We use \ours{}-L throughout, as it performs best in \Cref{exp:mof_sp}.

\textbf{Data preparation \& hyperparameters.}
We apply the preprocessing pipeline from \Cref{exp:mof_sp} to local-minimum configurations from the filtered ODAC25 trajectories, yielding 153{,}249 training and 3{,}582 validation datapoints. Since the official test set is unavailable, we use the validation split for testing and randomly hold out 3{,}130 training datapoints for validation. Hyperparameters and curriculum adjustments are detailed in \Cref{app:train_details}.

\subsection{Coverage of adsorption configurations} \label{exp:odac25_cov}

\textbf{Baselines.} We compare \ours{} with 10 ODE steps and no steering against grand canonical Monte Carlo (GCMC) and random sampling (RS). Since GCMC and RS only place adsorbates, we use the \ours{}-predicted framework and evaluate adsorbate placement accuracy. We run GCMC in \texttt{RASPA3}~\citep{ran2024raspa3} with classical force fields. For RS, we repeatedly insert a fixed number of adsorbates and reject steric clashes, defined as any pairwise distance $d_{ij}\le 0.8(r_i+r_j)$ using covalent radii $r_i$ and $r_j$. Implementation details and baseline discussion appear in \Cref{app:mof_ads_details}.

\textbf{Metrics.} We report \emph{coverage} (COV) \citep{wang2024swallowing}, which measures how many reference configurations are recovered for a given MOF-adsorbate system. For a given system, let $\mathcal{D}=\{\mathcal{S}^{(n)}\}_{n=1}^{N}$ be the set of ground-truth configurations and $\hat{\mathcal{D}}$ the set of predictions. Let $\hat{\mathcal{D}}_n \subseteq \hat{\mathcal{D}}$ denote the subset of predictions that match $\mathcal{S}^{(n)}$ under \texttt{StructureMatcher} with $\operatorname{stol}=0.5$. We then compute the fraction of recovered structures as $\operatorname{COV} = \frac{1}{N}\sum_{n} \mathbb{I}[|\hat{\mathcal{D}}_n| > 0]$. 

\textbf{Results.} \Cref{fig:coverage_vs_time} shows that \ours{} achieves higher coverage than both baselines under the same time budget, and is orders of magnitude faster than GCMC. \ours{} reaches $23.11\%$ coverage in under 1\,s, whereas GCMC requires nearly $10^{4}$\,s to reach $21.06\%$. RS runs on a time scale comparable to \ours{} but consistently attains lower coverage, indicating that \ours{} learns a distribution that better matches the reference adsorption configurations. See \Cref{app:adsorption_diversity} for a complementary diversity evaluation.

\subsection{Adsorption energy success rate} \label{exp:odac25_ads}

\definecolor{oursteal}{HTML}{00BFC4}

\begin{wrapfigure}{r}{\odacviswidth}
    \vspace{-15pt}
    \centering
    \small
    \resizebox{\odacvisscale\linewidth}{!}{%
    \begin{tabular}{cccc}
        \toprule
        {\Large \ce{H2O}}
        & {\Large \ce{CO2}}
        & {\Large \ce{N2}} \\
        \midrule
        \includegraphics[height=2.2cm]{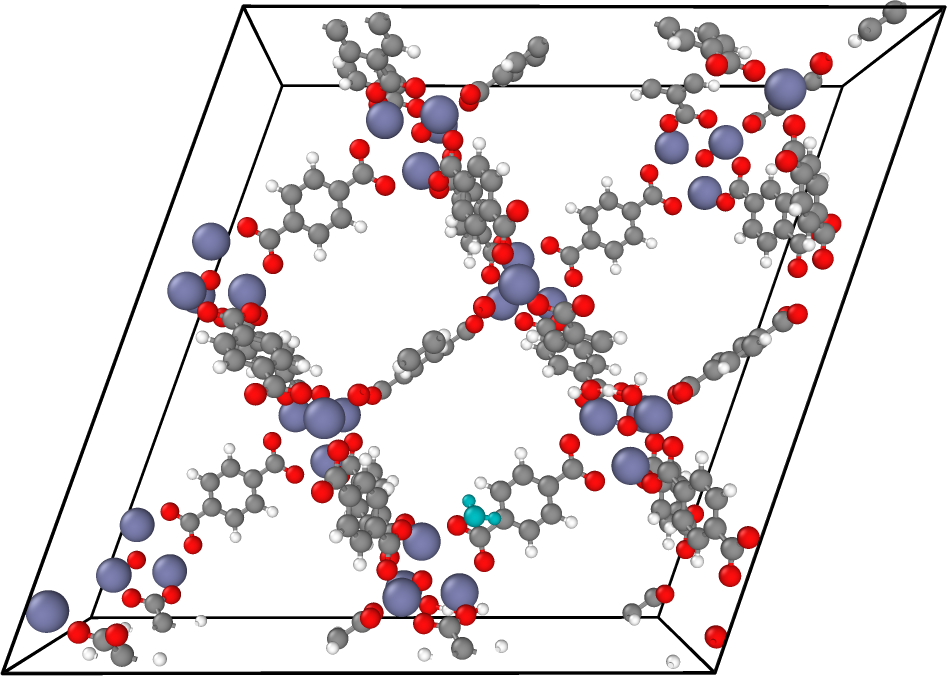} &
        \includegraphics[height=2.2cm]{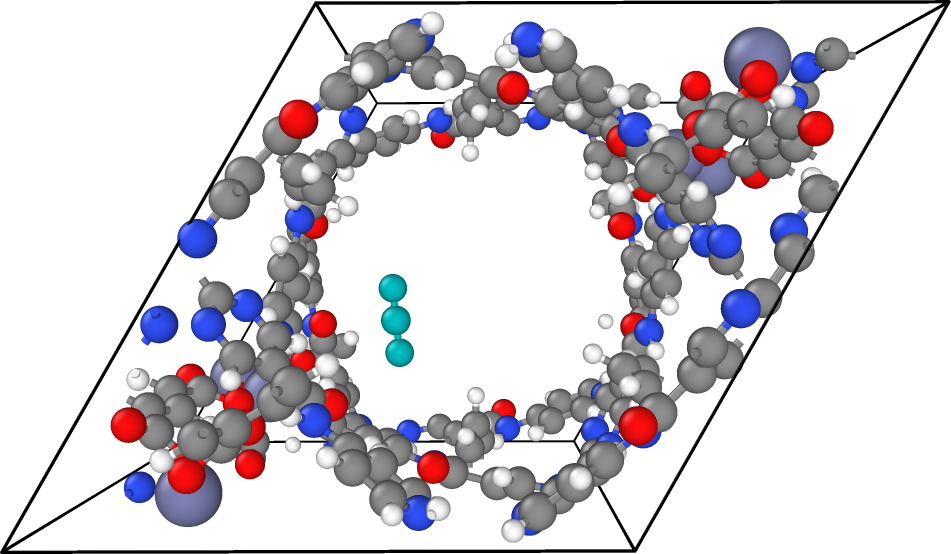} &
        \includegraphics[height=2.2cm]{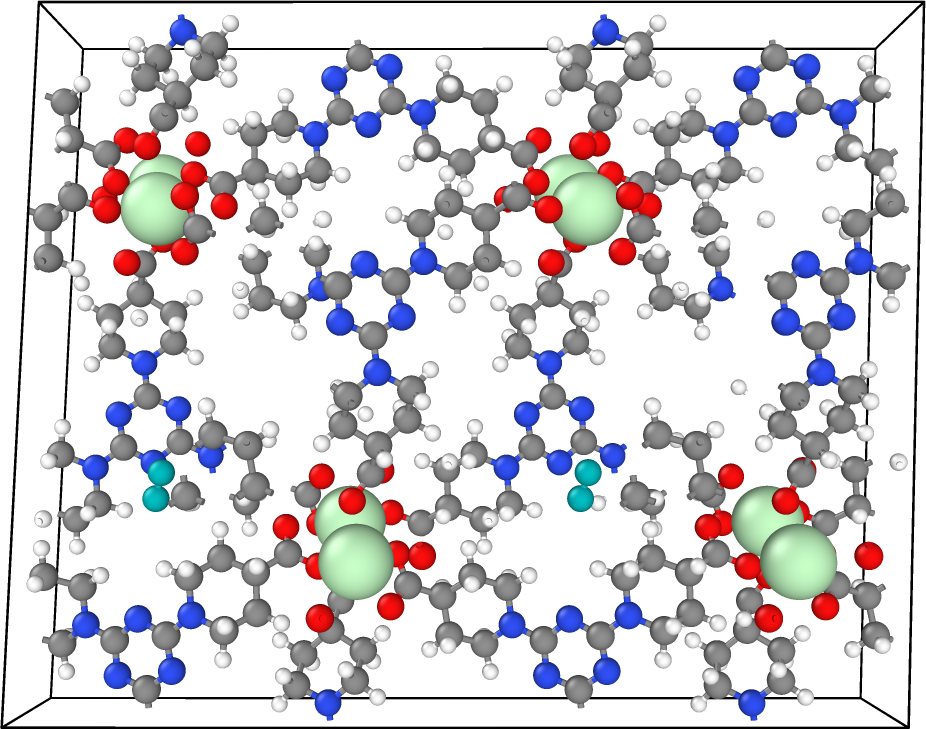} \\
        \bottomrule
    \end{tabular}%
    }
    \caption{\textbf{Examples.} MOF-adsorbate configurations predicted by \ours{}. Adsorbates are highlighted in \hlc{oursteal}{20}{teal} for clarity.}
    \label{fig:odac25_visualization}
    \vspace{-15pt}
\end{wrapfigure}

\textbf{Sampling and relaxation.} We generate 1 and 10 candidates per test datapoint. To improve stability during MLIP relaxation, we apply Feynman-Kac steering with the penalty $E_{\text{overlap}} = \sum_{i < j} \max(0, 0.8 (r_i + r_j) - d_{ij})^2$, where $d_{ij}$ is the interatomic distance and $r_i$ and $r_j$ are the covalent radii. We then filter remaining overlapping structures using the criterion $d_{ij} \le 0.8(r_i + r_j)$ and relax structures with \texttt{eSEN-ODAC25-Full}.

\textbf{Metrics.} Following \citet{lan2023adsorbml} and \citet{kolluru2024adsorbdiff}, we report success rate. A prediction is successful if: (1) MLIP relaxation converges within 300 steps to a maximum force below $0.05\text{ \eva}$ and (2) $\varepsilon_{\min} \le 0.1\text{ eV}$, where $\varepsilon_{\min} = \min_{\hat{\mathcal{S}} \in \hat{\mathcal{D}}} \Delta E_{\text{ads}}(\hat{\mathcal{S}}) - \min_{\mathcal{S} \in \mathcal{D}} \Delta E_{\text{ads}}(\mathcal{S})$. Because DFT relaxation and energy evaluation are prohibitively expensive for large MOF-adsorbate systems, we use MLIP-based adsorption energies as a surrogate, following prior work~\citep{simkus2025mofasa, liu2026promof}. To use a different MLIP for evaluation than for relaxation, we use UMA (\texttt{uma-s-1p2}).

\begin{wraptable}{r}{\successtablewidth}
    \vspace{\successtabletopskip}
    \caption{\textbf{Adsorption energy success rate.} Success rate and minimum energy difference ($\varepsilon_{\min}$) under UMA (\texttt{uma-s-1p2}) evaluation for 1 and 10 samples. Structures are relaxed with eSEN; average $\varepsilon_{\min}$ is computed over converged cases.}
    \label{tab:odac25_success_rate}
    \centering
    \resizebox{\successtablewidth}{!}{%
    \begin{tabular}{lrr}
    \toprule
    Metric & 1 Sample & 10 Samples \\
    \midrule
    Success rate (\%) & 13.18 & 20.64 \\
    $\varepsilon_{\min} < 0$ (\%) & 6.59 & 16.40 \\
    Avg. $\varepsilon_{\min}$ (eV) & 0.535 & 0.039 \\
    \bottomrule
    \end{tabular}%
    }
    \vspace{-1.5em}
\end{wraptable}

\textbf{Results.} \Cref{tab:odac25_success_rate} shows that \ours{} achieves success rates of 13.18\% with 1 sample and 20.64\% with 10 samples. It also finds configurations with lower adsorption energies than the reference ($\varepsilon_{\min} < 0$) in 6.59\% and 16.40\% of systems, respectively, which is important for accurate adsorption-energy estimation~\citep{lan2023adsorbml}. With 10 samples, the average minimum energy difference is 0.039~eV, indicating that the best configurations are close to the reference minimum. This agrees with \Cref{fig:ads_energy_hist}, where the distribution of minimum adsorption energies closely matches the reference. Representative adsorption configurations are shown in \Cref{fig:odac25_visualization}.

\section{Conclusion} \label{conclusion}

We present \ours{}, a scalable flow-based model for MOF-adsorbate structure prediction. Unlike prior approaches with structural constraints, \ours{} models all atoms directly, which we find necessary for high structural accuracy. We also show predictable scaling with model size and compute. To improve sampling quality, we introduce MLIP-guided Feynman-Kac steering, which increases structural stability and validity. Empirically, \ours{} outperforms prior baselines on BW. On ODAC25, it recovers diverse adsorption configurations far faster than GCMC and provides MLIP relaxation initializations that can reach lower adsorption energies than the reference.

\textbf{Limitations \& future work.}
\ours{} is less accurate on very large and out-of-distribution systems, such as those with unseen metal nodes or adsorbates, and it complements rather than replaces GCMC, which also yields thermodynamic quantities such as Henry constants and adsorption isotherms. Moreover, match rate cannot credit valid polymorphs absent from the references, although a MOFid~\citep{bucior2019identification} analysis found no such topological diversity in the BW test set. Future work includes broader data coverage, thermodynamic prediction, and polymorph-aware evaluation.

\section*{Acknowledgements}
This work was supported by
Artificial intelligence industrial convergence cluster development project funded by the Ministry of Science and ICT(MSIT, Korea) \& Gwangju Metropolitan City; 
Advanced GPU Utilization Support Program(Beta Service) funded by the Government of the Republic of Korea (Ministry of Science and ICT); 
Advanced GPU Utilization Support Program funded by the Government of the Republic of Korea (Ministry of Science and ICT); 
Basic Science Research Program through the National Research Foundation of Korea(NRF) funded by the Ministry of Education (RS-2025-25435147); 
Institute for Information \& communications Technology Planning \& Evaluation(IITP) grant funded by the Korea government(MSIT) (RS-2019-II190075, Artificial Intelligence Graduate School Support Program(KAIST)); 
National Research Foundation of Korea(NRF) grant funded by the Korea government(MSIT) (RS-2025-02216257);
GRDC(Global Research Development Center) Cooperative Hub Program through the National Research Foundation of Korea(NRF) grant funded by the Ministry of Science and ICT(MSIT) (No. RS-2024-00436165); 
and Institute of Information \& Communications Technology Planning \& Evaluation(IITP) grant funded by the Korea government(MSIT) (RS-2025-02304967, AI Star Fellowship(KAIST)). 

\bibliographystyle{abbrvnat}
\bibliography{ref}

\clearpage
\begingroup
\setcounter{tocdepth}{2}
\setlength{\parskip}{0pt}
\setlength{\fboxsep}{14pt}
\makeatletter
\let\appendixcontentsline\contentsline
\renewcommand{\contentsline}[4]{}
\renewcommand{\appendixcontentsstart}{\let\contentsline\appendixcontentsline}
\renewcommand{\l@section}[2]{%
  \addvspace{0.6em}%
  {\bfseries\def\@dotsep{10000}\@dottedtocline{1}{0em}{1.8em}{#1}{\bfseries #2}}}
\renewcommand{\l@subsection}{\@dottedtocline{2}{0em}{2.8em}}
\noindent\fbox{%
  \begin{minipage}{\dimexpr\linewidth-2\fboxsep-2\fboxrule\relax}
    \begin{center}\LARGE\textsc{Appendix}\end{center}
    \vspace{4pt}
    \@starttoc{toc}
  \end{minipage}%
}
\makeatother
\endgroup
\clearpage
\appendix
\addtocontents{toc}{\protect\appendixcontentsstart}
\crefalias{section}{appendix}
\crefalias{subsection}{subappendix}
\crefalias{subsubsection}{subsubappendix}
\raggedbottom
\section{Model architecture} \label{app:architecture}

Here we describe the detailed implementation of our denoiser, $\boldsymbol{\mu}_t^{\theta}(\mathbf{X}_t, \boldsymbol{\ell}_t \mid \mathcal{B})$ (\Cref{alg:flow_module}). As shown in \Cref{fig:model_architecture}, the architecture consists of three stages: (i) an \textit{encoder} that featurizes the building blocks $\mathcal{B}$ into single-atom and pairwise representations (\Cref{alg:encoder}); (ii) a \textit{trunk} that performs the main computation over atomic interactions (\Cref{alg:trunk}); and (iii) a \textit{decoder} that maps the transformed features back to per-atom outputs to predict $\hat{\mathbf{X}}_1$ and $\hat{\boldsymbol{\ell}}_1$ (\Cref{alg:decoder}).

\begin{figure}[H]
    \centering
    \includegraphics[width=\linewidth]{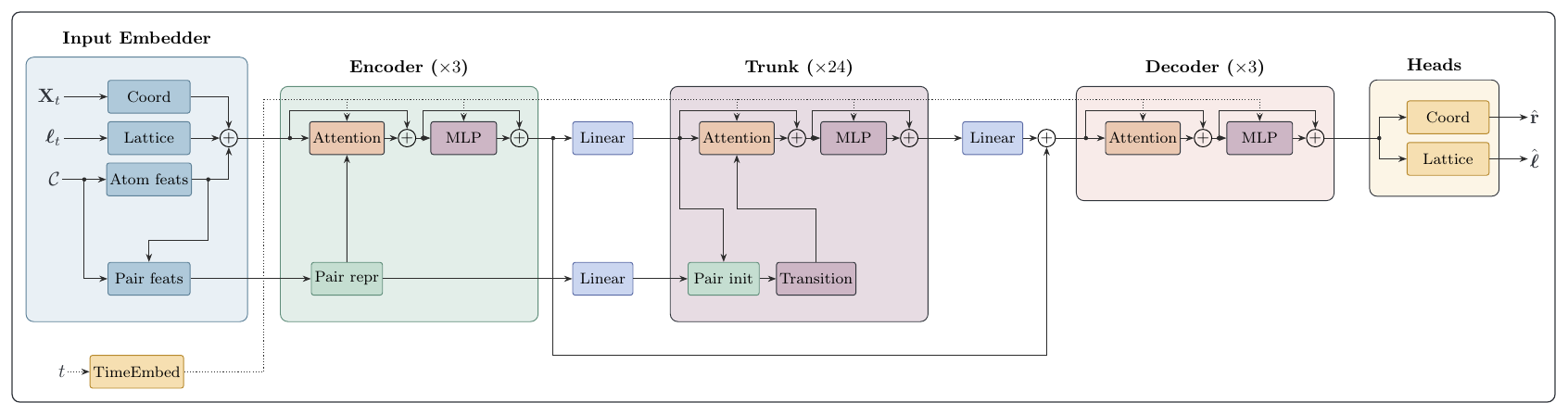}
    \caption{\textbf{AtomMOF architecture.} Building-block features and noisy coordinates and lattice parameters are processed by an encoder, trunk, and decoder to predict the clean structure. Pairwise representations condition attention, while timestep embeddings condition the network blocks.}
    \label{fig:model_architecture}
\end{figure}

\begin{algorithm}[H]
   \caption{Denoiser $\boldsymbol{\mu}_t^{\theta}(\mathbf{X}_t, \boldsymbol{\ell}_t \mid \mathcal{B})$  }
   \label{alg:flow_module}
\begin{algorithmic}
   \State {\bfseries Input:} Noisy inputs $\bm{X}_t, \bm{\ell}_t$, Time $t$, building block $\mathcal{B}$.
   \State {\bfseries Output:} Denoised updates $\hat{\mathbf{X}}_1, \hat{\bm{\ell}}_1$.

   \State \textit{\# 1. Encoding}
   \State $\bm{Q}, \bm{P} \leftarrow \text{Encoder}(\bm{X}_t, \bm{\ell}_t, t, \mathcal{B})$

   \State \textit{\# 2. Projection}
   \State $\bm{S} \leftarrow \text{Linear}(\bm{Q})$
   \State $\bm{Z} \leftarrow \text{Linear}(\bm{P})$

   \State \textit{\# 3. Trunk}
   \State $\bm{S}' \leftarrow \text{Trunk}(\bm{S}, \bm{Z}, t)$

   \State \textit{\# 4. Residual projection}
   \State $\bm{Q} \leftarrow \bm{Q} + \text{Linear}(\bm{S}')$

   \State \textit{\# 5. Decoding}
   \State $\hat{\mathbf{X}}_1, \hat{\bm{\ell}}_1 \leftarrow \text{Decoder}(\bm{Q}, t)$

   \State {\bfseries Return:} $\hat{\mathbf{X}}_1, \hat{\bm{\ell}}_1$
\end{algorithmic}
\end{algorithm}

\begin{algorithm}[H]
   \caption{Encoder}
   \label{alg:encoder}
\begin{algorithmic}
   \State {\bfseries Input:} Noisy coordinates $\bm{X}_t$, noisy lattice $\bm{\ell}_t$, time $t$, building block $\mathcal{B}$.
   \State {\bfseries Output:} Single representation $\bm{Q}$, pair representation $\bm{P}$.

   \State \textit{\# 1. Feature embedding}
   \State Concatenate $\mathcal{B}$\textquotesingle s RDKit-initialized $\bm{X}_\text{local}$, formal charges, and atom types into $\bm{f}_{\text{atom}}$.
   \State Initialize single representation: $\bm{C} \leftarrow \text{LinearNoBias}(\bm{f}_{\text{atom}})$

   \State \textit{\# 2. Pairwise representation construction}
   \State Compute pairwise distances $\bm{D}_{ij} \leftarrow \bm{X}_{\text{local},i} - \bm{X}_{\text{local},j}$ and norms $\|\bm{D}_{ij}\|$.
   \State Construct block adjacency mask $\bm{v}_{ij}$.
   \State Initialize pair representation $\bm{P}$:
   \State $\bm{P} \leftarrow \text{Embed}_{\text{pos}}(\bm{D}) + \text{Embed}_{\text{dist}}(\|\bm{D}\|) + \text{Embed}_{\text{bond}}(\mathcal{B}_{\text{bond}})$.
   \State $\bm{P} \leftarrow \bm{P} \odot \bm{v} + \text{Embed}_{\text{mask}}(\bm{v}) \odot \bm{v}$.

   \State \textit{\# 3. Condition integration}
   \State Add single context to pairs:
   \State $\bm{P}_{ij} \leftarrow \bm{P}_{ij} + \text{MLP}_Q(\bm{C}_i) + \text{MLP}_K(\bm{C}_j) + \text{MLP}_{\text{pair}}(\bm{P}_{ij})$.
   \State Add noise conditioning to single rep $\bm{Q}$:
   \State $\bm{Q} \leftarrow \bm{C} + \text{Linear}(\bm{X}_t) + \text{Linear}(\bm{\ell}_t)$.

   \State \textit{\# 4. Encoding}
   \State $\bm{Q}, \bm{P} \leftarrow \text{DiT}_{\text{encoder}}(\bm{Q}, \bm{P}, t)$.

   \State {\bfseries Return:} $\bm{Q}, \bm{P}$
\end{algorithmic}
\end{algorithm}

\begin{algorithm}[H]
   \caption{Trunk}
   \label{alg:trunk}
\begin{algorithmic}
   \State {\bfseries Input:} Single representation $\bm{S}_{\text{in}}$, pair representation $\bm{Z}_{\text{in}}$, time $t$.
   \State {\bfseries Output:} Updated single representation $\bm{S}'$.

   \State \textit{\# 1. Initialization}
   \State $\bm{S} \leftarrow \text{Linear}_{\text{init}}(\bm{S})$
   \State $\bm{Z}_{\text{bias}} \leftarrow \text{Linear}_1(\bm{S}) \oplus \text{Linear}_2(\bm{S})^T$
   \State $\bm{Z} \leftarrow \bm{Z} + \bm{Z}_{\text{bias}}$

   \State \textit{\# 2. Normalization}
   \State $\bm{S} \leftarrow \text{Linear}(\text{LayerNorm}(\bm{S}))$,  $\bm{Z} \leftarrow \text{Linear}(\text{LayerNorm}(\bm{Z}))$

   \State \textit{\# 3. Transformer Pass}
   \State $\bm{S}_{\text{out}} \leftarrow \text{DiT}_{\text{trunk}}(\bm{S}, \bm{Z}, t)$

   \State {\bfseries Return:} $\bm{S}'$
\end{algorithmic}
\end{algorithm}

\begin{algorithm}[H]
   \caption{Decoder}
   \label{alg:decoder}
\begin{algorithmic}
   \State {\bfseries Input:} Single representation $\bm{Q}$, time $t$.
   \State {\bfseries Output:} Coordinate prediction $\hat{\mathbf{X}}_1$, lattice prediction $\hat{\bm{\ell}}_1$.

   \State \textit{\# 1. Feature decoding}
   \State $\bm{Q}_{\text{dec}} \leftarrow \text{DiT}_{\text{decoder}}(\bm{Q}, t)$

   \State \textit{\# 2. Coordinate projection}
   \State $\hat{\mathbf{X}}_1 \leftarrow \text{Linear}(\text{LayerNorm}(\bm{Q}_{\text{dec}}))$

   \State $\bm{Q}_{\text{global}} \leftarrow{} \text{Average} (\bm{Q}_{\text{dec}})$
   \State $\hat{\bm{\ell}}_1 \leftarrow \text{Linear}(\text{LayerNorm}(\bm{Q}_{\text{global}}))$

   \State {\bfseries Return:} $\hat{\mathbf{X}}_1, \hat{\bm{\ell}}_1$
\end{algorithmic}
\end{algorithm}

\begin{algorithm}[H]
   \caption{Diffusion Transformer (DiT) Architecture}
   \label{alg:dit}
\begin{algorithmic}
   \State {\bfseries Input:} Input $\bm{S}$, time $t$, (optional) bias $\bm{Z}$. 
   \State {\bfseries Output:} Updated $\bm{S}$.

   \State \textit{\# 1. Time conditioning}
   \State $\bm{c} \leftarrow \text{MLP}(\text{SinusoidalEmbedding}(t))$

   \For{$l = 1 \dots L$}
      \State \textit{\# 2. Adaptive modulation (AdaLN)}
      \State $\bm{\gamma}_{1,2}, \bm{\beta}_{1,2}, \bm{\alpha}_{1,2} \leftarrow \text{Chunk}(\text{Linear}(\text{SiLU}(\bm{c})), 6)$

      \State \textit{\# 3. Self-attention block}
      \State $\bm{S}_{\text{norm}} \leftarrow \text{LayerNorm}(\bm{S}) \cdot (1 + \bm{\gamma}_1) + \bm{\beta}_1$
      \State $\bm{S} \leftarrow \bm{S} + \bm{\alpha}_1 \cdot \text{Attn}(\bm{S}_{\text{norm}}, \text{bias}=\bm{Z})$

      \State \textit{\# 4. Feed-forward block}
      \State $\bm{S}_{\text{norm}} \leftarrow \text{LayerNorm}(\bm{S}) \cdot (1 + \bm{\gamma}_2) + \bm{\beta}_2$
      \State $\bm{S} \leftarrow \bm{S} + \bm{\alpha}_2 \cdot \text{MLP}(\bm{S}_{\text{norm}})$
   \EndFor

   \State {\bfseries Return:} $\bm{S}$
\end{algorithmic}
\end{algorithm}

\begin{algorithm}[H]
   \caption{Attention with pairwise bias}
   \label{alg:attention}
\begin{algorithmic}
   \State {\bfseries Input:} Input $\bm{X}$, (optional) pairwise bias $\bm{Z}$.
   \State {\bfseries Output:} Attention output $\bm{Y}$

   \State $\mathsf{Q}, \mathsf{K}, \mathsf{V} \leftarrow \text{Linear}_{q,k,v}(\bm{X})$

   \State $\bm{A} \leftarrow \operatorname{softmax}\left( \frac{\mathsf{Q}\mathsf{K}^{\top}}{\sqrt{D_k}} + \operatorname{Linear}_{\text{bias}}(\bm{Z}) \right) \mathsf{V}$

   \State $\bm{Y} \leftarrow \text{Linear}_{\text{out}}(\bm{A})$

   \State {\bfseries Return:} $\bm{Y}$
\end{algorithmic}
\end{algorithm}

\flushbottom
\clearpage
\section{MLIP-guided Feynman-Kac steering algorithm} \label{app:fk_algorithm}

In this section, we present the details of the MLIP-guided Feynman–Kac (FK) steering algorithm. The key idea is to bias the sampling trajectory towards low-energy structures by maintaining an ensemble of $K$ particles, updating their weights at intermediate steps using a potential, and resampling the particles according to these weights. As summarized in \Cref{tab:potentials}, the potential can depend on the current state, consecutive states, or the trajectory history. Algorithm~\ref{alg:fk_steering} details the full FK procedure.

\begin{table}[h]
    \caption{\textbf{Overview of potential functions.} Here, $\mathcal{E}_t \coloneqq E(\hat{\mathcal{S}}_1)$ denotes the potential energy of the clean state predicted at time $t$. $\Delta t$ refers to the resampling interval and $N_t$ denotes the number of steps elapsed up to time $t$.}
    \label{tab:potentials}
    \centering
    \small
    \begin{tabular}{lll}
        \toprule
        \textbf{Potential} & \textbf{Domain} & \textbf{Definition} $G_t(\cdot)$ \\
        \midrule
        Difference & $\mathcal{S}_{t-\Delta t}, \mathcal{S}_t$ & $\exp\left(\lambda (\mathcal{E}_{t-\Delta t} - \mathcal{E}_{t})\right)$ \\
        \addlinespace
        Immediate  & $\mathcal{S}_t$ & $\exp\left(-\lambda \mathcal{E}_t\right)$ \\
        \addlinespace
        Max        & $\{\mathcal{S}_{\tau}\}_{\tau\in[0,t]}$ & $\exp\left(- \lambda \min_{\tau \le t} \mathcal{E}_\tau\right)$ \\
        \addlinespace
        Sum        & $\{\mathcal{S}_{\tau}\}_{\tau\in[0,t]}$ & $\exp(-\tfrac{\lambda}{N_t} \sum_{\tau \le t} \mathcal{E}_\tau)$ \\
        \bottomrule
    \end{tabular}%
\end{table}
\begin{algorithm}[H]
   \caption{Feynman--Kac Steering with MLIP}
   \label{alg:fk_steering}
\begin{algorithmic}
   \State {\bfseries Input:} Denoiser $\boldsymbol{\mu}_\theta$, MLIP $E(\cdot)$.
   \State {\bfseries Hyperparams:} Steering strength $\lambda$, particles $K$, resampling interval $\Delta \tau$.
   \State {\bfseries Initialize:} Sample $\mathcal{S}_0^{(k)} \sim p_0$ for $k=1, \dots, K$.
   \State {\bfseries Initialize:} History of energies $\mathcal{E}_{\text{hist}}^{(k)} \leftarrow \emptyset$ for all $k$.
   \State Define time discretization $\{t_i\}_{i=0}^T$ with $t_0=0, t_T=1$.
   
   \For{$i=0$ {\bfseries to} $T-1$}
       \State $t \leftarrow t_i$, $\Delta t \leftarrow t_{i+1} - t_i$.
       \State \textit{\# 1. Predict clean state}
       \State $\hat{\mathcal{S}}_1^{(k)} \leftarrow \boldsymbol{\mu}_t^\theta(\bm{X}_t^{(k)}, \bm{\ell}_t^{(k)} \mid \mathcal{B})$ for all $k$.
       \State Compute current energies: $\mathcal{E}_{\text{curr}}^{(k)} \leftarrow E(\hat{\mathcal{S}}_1^{(k)})$.
       
       \State \textit{\# 2. Resample (Feynman--Kac)}
       \If{$t \geq t_{\text{start}}$ \textbf{and} $i \bmod \Delta \tau = 0$}
           \State Compute weights $w_k$ using $G_t$ (see \Cref{tab:potentials}).
           \State \textit{e.g., for difference:} $w_k \leftarrow \exp(\lambda(\mathcal{E}_{t - \Delta t}^{(k)} - \mathcal{E}_{t}^{(k)}))$.
           \State Normalize weights: $\tilde{w}_k \leftarrow w_k / \sum_{j=1}^K w_j$.
           \State Resample indices $\{j_k\}_{k=1}^K \sim \text{Multinomial}(\{\tilde{w}_k\})$.
           \State Update particles: $\mathcal{S}_t^{(k)} \leftarrow \mathcal{S}_t^{(j_k)}$, $\mathcal{E}_{\text{curr}}^{(k)} \leftarrow \mathcal{E}_{\text{curr}}^{(j_k)}$.
       \EndIf
       \State Update history: $\mathcal{E}_{\text{hist}}^{(k)} \leftarrow (\mathcal{E}_{\text{hist}}^{(k)}, \mathcal{E}_{\text{curr}}^{(k)})$

       \State \textit{\# 3. SDE update (for all $k$)}
       \State \textit{\# Step lattice:}
       \State $\mathbf{v}_{t,\bm{\ell}}^{(k)} \leftarrow (\hat{\bm{\ell}}_1^{(k)} - \bm{\ell}_t^{(k)}) / (1-t)$
       \State $\bm{\ell}_{t+\Delta t}^{(k)} \leftarrow \bm{\ell}_t^{(k)} + \mathbf{v}_{t, \bm{\ell}}^{(k)} \Delta t$
       \State \textit{\# Step coordinates:}
       \State $\mathbf{v}_{t,\bm{X}}^{(k)} \leftarrow (\hat{\bm{X}}_1^{(k)} - \bm{X}_t^{(k)}) / (1-t)$
       \State $s_t^{(k)} \leftarrow (t \mathbf{v}_{t,\bm{X}}^{(k)} - \bm{X}_t^{(k)}) / (1-t)$, sample $\bm{z}^{(k)} \sim \mathcal{N}_C(\bm{0}, \bm{I})$, compute $g(t) \leftarrow \sqrt{(1-t)/2}$
       \State $\bm{X}_{t+\Delta t}^{(k)} \leftarrow \bm{X}_t^{(k)} + (\mathbf{v}_{t, \bm{X}}^{(k)} + \tfrac{1}{2} g(t)^2 s_t^{(k)}) \Delta t + g(t) \sqrt{\Delta t}\,\bm{z}^{(k)} $
   \EndFor
   \State {\bfseries Return:} $\{\mathcal{S}_1^{(k)}\}_{k=1}^K$
\end{algorithmic}
\end{algorithm}
\clearpage
\section{Training details} \label{app:train_details}

\textbf{Metal node template construction.} For the BW dataset, node-template construction is simple because the dataset contains only 9 recurring node types: for each type, we align all training-set instances with the Kabsch algorithm and average the aligned coordinates, yielding one canonical template per node type. For ODAC25, this averaging strategy is not appropriate because metal-node geometries are substantially more diverse. Instead, we keep the training set geometries themselves and build a dictionary from node SMILES to all matching node structures, covering 883 unique nodes. At inference time, we use this dictionary only to initialize node coordinates by randomly sampling one training-set geometry with the same SMILES. Nearly all validation nodes are covered by this dictionary (472/473), and the rare missing cases are initialized from a Gaussian distribution.

\textbf{Training setup.} We train all model variants with dynamic batching based on a squared-atom limit ($N_{\text{max}}^2$) to handle the high variability in atomic counts. We optimize the model with AdamW and a linear warmup scheduler, and improve convergence and stability with a multi-stage curriculum that progressively increases the maximum number of atoms used in each stage; this curriculum is shared across \ours{}-S, \ours{}-M, and \ours{}-L. Architectural hyperparameters are listed in \Cref{tab:hyperparams}, optimization settings in Table~\ref{tab:training_params}, and the curriculum schedule in \Cref{tab:curriculum}.

For ODAC25, we reuse the BW hyperparameters, increasing the curriculum atom-count limit by 100 atoms to improve dataset coverage.

\textbf{Compute resources.}
Training used distributed data parallelism on 8 GPUs. AtomMOF-S and AtomMOF-L were trained on NVIDIA B200 GPUs with approximately 180~GB memory per GPU, while AtomMOF-M was trained on NVIDIA RTX Pro 6000 Blackwell Max-Q GPUs with 96~GB memory per GPU. Table~\ref{tab:compute_resources} reports the wall-clock time and GPU-hour estimate for each primary training run; reproducing the released primary checkpoints requires approximately 3,490 GPU-hours in total, excluding preprocessing, inference/evaluation, and exploratory or failed runs. 

\textbf{License.} This work uses the BW~\citep{boyd2019data} and ODAC25~\citep{sriram2025open} datasets. BW uses a Creative Commons Attribution license; ODAC25 uses the FAIR Chemistry License v1, which permits use, redistribution, and derivative works subject to the license terms, attribution requirements, applicable law, and the FAIR Chemistry Acceptable Use Policy.

\textbf{Softwares.} Existing software assets used in this work include RDKit~\citep{rdkit}, pymatgen~\citep{ong2013pymatgen}, Zeo++~\citep{willems2012algorithms}, RASPA3~\citep{ran2024raspa3}, ASE~\citep{larsen2017atomic}, and OVITO~\citep{ovito}.

\begin{table}[h]
\centering
\small
\caption{Hyperparameters for \ours{}-S, \ours{}-M, and \ours{}-L.}
\label{tab:hyperparams}
\begin{tabular}{lccc}
\toprule
\textbf{Parameter} & \textbf{\ours{}-S} & \textbf{\ours{}-M} & \textbf{\ours{}-L} \\
\midrule
\multicolumn{4}{l}{\textbf{\textit{Encoder \& Decoder}}} \\
\quad Single dim. & 256 & 256 & 256 \\
\quad Pairwise dim. & 128 & 128 & 128 \\
\quad Depth & 3 & 3 & 3 \\
\quad Heads & 4 & 4 & 4 \\
\addlinespace
\multicolumn{4}{l}{\textbf{\textit{Trunk}}} \\
\quad Single dim. & 256 & 512 & 1024 \\
\quad Pairwise dim. & 128 & 256 & 512 \\
\quad Depth & 24 & 24 & 24 \\
\quad Heads & 8 & 8 & 16 \\
\midrule
\textbf{Total params.} & 38.4M & 129M & 491M \\
\midrule
\multicolumn{4}{l}{\textbf{\textit{Batch limit ($N_{\text{max}}^2$)}}} \\
\quad Pretraining & $3.0 \times 10^6$ & $1.0 \times 10^6$ & $1.0 \times 10^6$ \\
\quad Fine-tuning 1 & $3.0 \times 10^6$ & $1.0 \times 10^6$ & $1.0 \times 10^6$ \\
\quad Fine-tuning 2 & $2.5 \times 10^6$ & $8.0 \times 10^5$ & $8.0 \times 10^5$ \\
\bottomrule
\end{tabular}
\end{table}

\begin{table}[h]
\centering
\small
\caption{Training and optimization hyperparameters.}
\label{tab:training_params}
\begin{tabular}{lc}
\toprule
\textbf{Parameter} & \textbf{Value} \\
\midrule
\multicolumn{2}{l}{\textbf{\textit{Loss weights}}} \\
\quad Coordinate weight & 1.0 \\
\quad Lengths weight & 1.0 \\
\quad Angle weight & 0.01 \\
\addlinespace
\multicolumn{2}{l}{\textbf{\textit{Optimizer (AdamW)}}} \\
\quad Learning rate & $1 \times 10^{-4}$ \\
\quad Weight decay & 0.0 \\
\addlinespace
\multicolumn{2}{l}{\textbf{\textit{Scheduler (linear warmup)}}} \\
\quad Base LR & 0.0 \\
\quad Warmup steps & 10,000 \\
\bottomrule
\end{tabular}
\end{table}

\begin{table}[h]
\centering
\small
\caption{\textbf{Curriculum learning schedule.} Maximum atom count and training epochs for each stage, shared across \ours{}-S, \ours{}-M, and \ours{}-L.}
\label{tab:curriculum}
\begin{tabular}{lccc}
\toprule
\textbf{Setting} & \textbf{Pretraining} & \textbf{Fine-tuning 1} & \textbf{Fine-tuning 2} \\
\midrule
\multicolumn{4}{l}{\textbf{\textit{Maximum atom count}}} \\
\quad BW & 200 & 300 & No limit \\
\quad ODAC25  & 300 & 400 & No limit \\
\midrule
\textbf{Epochs} & 500 & 100 & 500 \\
\bottomrule
\end{tabular}
\end{table}

\begin{table}[t]
\centering
\small
\caption{Compute resources for primary \ours{} training runs. GPU-hours are computed as wall-clock hours $\times$ 8 GPUs.}
\label{tab:compute_resources}
\begin{tabular}{llrr}
\toprule
Experiment & Stage & Wall-clock & GPU-hours \\
\midrule
\ours{}-S (BW) & Pretraining & 23.0 h & 184 \\
 & Fine-tuning 1 & 7.2 h & 58 \\
 & Fine-tuning 2 & 22.3 h & 178 \\
 & Total & 52.5 h & 420 \\
\midrule
\ours{}-M (BW) & Pretraining & 26.7 h & 214 \\
 & Fine-tuning 1 & 8.3 h & 66 \\
 & Fine-tuning 2 & 83.6 h & 669 \\
 & Total & 118.6 h & 949 \\
\midrule
\ours{}-L (BW) & Pretraining & 35.5 h & 284 \\
 & Fine-tuning 1 & 11.5 h & 92 \\
 & Fine-tuning 2 & 45.0 h & 360 \\
 & Total & 92.0 h & 736 \\
\midrule
\ours{}-L (ODAC25) & Pretraining & 54.3 h & 434 \\
 & Fine-tuning 1 & 19.1 h & 153 \\
 & Fine-tuning 2 & 99.8 h & 798 \\
 & Total & 173.1 h & 1,385 \\
\bottomrule
\end{tabular}
\end{table}
\clearpage
\section{Additional ablations and controls} \label{app:ablation_details}

We evaluate building-block features, curriculum training, and deterministic prediction using \ours{}-M on BW at $\operatorname{stol}=0.3$ and 0.5.

\textbf{Model architecture ablations.}
We ablate pairwise representations and bond topology, training each variant for 500 epochs on structures with at most 200 atoms (83.5\% of training data). Removing pairwise representations reduces match rate by 25.8\% at $\operatorname{stol}=0.3$ (\Cref{tab:ablation_features}). Removing bond topology reduces it by only 7.9\%, suggesting that atom types and pairwise distances retain some connectivity information.

\textbf{Curriculum training.}
We retrain \ours{}-M without curriculum on the full dataset for a comparable number of steps ($\sim$863K). Curriculum improves all reported metrics (\Cref{tab:ablation_curriculum}), but the model without it still outperforms MOFFlow-2, suggesting that the architecture contributes most.

\begin{table}[h]
    \centering
    \small
    \caption{\textbf{Ablation on building block feature initialization.} Removing the pairwise representation degrades performance most, while bond topology has a smaller effect.}
    \label{tab:ablation_features}
    \setlength{\tabcolsep}{6pt}
    \begin{tabular}{l c c}
        \toprule
        & \multicolumn{2}{c}{\textbf{Match rate (\%)} $\uparrow$} \\
        \cmidrule(lr){2-3}
        \textbf{Configuration} & $\operatorname{stol}=0.3$ & $\operatorname{stol}=0.5$ \\
        \midrule
        \ours{}-M         & 11.18 & 30.80 \\
        \arrayrulecolor{black!30}\midrule
        \arrayrulecolor{black}
        $-$ Pairwise representation & \phantom{0}8.29  & 26.60 \\
        $-$ Bond topology & 10.30 & 30.07 \\
        \bottomrule
    \end{tabular}%
\end{table}

\begin{table}[h]
    \centering
    \small
    \caption{\textbf{Ablation on curriculum training.} Curriculum learning improves match rate and reduces RMSD at both tolerances.}
    \label{tab:ablation_curriculum}
    \setlength{\tabcolsep}{7pt}
    \begin{tabular}{c c c c c}
        \toprule
        & \multicolumn{2}{c}{$\operatorname{stol}=0.3$} & \multicolumn{2}{c}{$\operatorname{stol}=0.5$} \\
        \cmidrule(lr){2-3} \cmidrule(lr){4-5}
        \textbf{Configuration} & \textbf{MR (\%)} $\uparrow$ & \textbf{RMSD} $\downarrow$ & \textbf{MR (\%)} $\uparrow$ & \textbf{RMSD} $\downarrow$ \\
        \midrule
        Without curriculum & 12.88 & 0.1360 & 36.43 & 0.2615 \\
        With curriculum    & 14.33 & 0.1308 & 37.88 & 0.2542 \\
        \bottomrule
    \end{tabular}%
\end{table}

\subsection{Deterministic versus generative prediction}
\label{app:deterministic}

We compare generative and deterministic \ours{}-M variants with the same architecture, optimizer, learning-rate warmup, and batching strategy. The deterministic model directly regresses coordinates and lattice parameters in one step. It uses canonical building-block coordinates because random orientation augmentation prevented accurate prediction in our experiments; the generative model retains augmentation and iterative sampling.

\Cref{tab:deterministic} supports the generative formulation in two respects: it achieves higher single-sample accuracy at both tolerances, and additional sampling further improves match rate and RMSD, whereas repeated deterministic predictions remain identical. A third advantage is compatibility with inference-time steering: FK steering improves physical properties without retraining the model (\Cref{exp:feynman_kac}).

\begin{table}[ht]
\centering
\small
\caption{\textbf{Deterministic versus generative prediction on BW.} Both models use the \ours{}-M architecture. Deterministic predictions are identical across repeated evaluations. RMSD is computed over successfully matched structures.}
\label{tab:deterministic}
\begin{tabular}{llcccc}
\toprule
$\operatorname{stol}$ & Model & MR@1 (\%) $\uparrow$ & RMSD@1 (\AA) $\downarrow$ & MR@5 (\%) $\uparrow$ & RMSD@5 (\AA) $\downarrow$ \\
\midrule
0.3 & Deterministic & 4.00 & 0.1709 & 4.00 & 0.1709 \\
 & Generative & \textbf{11.18} & \textbf{0.1347} & \textbf{21.78} & \textbf{0.1274} \\
\midrule
0.5 & Deterministic & 17.74 & 0.3047 & 17.74 & 0.3047 \\
 & Generative & \textbf{30.80} & \textbf{0.2622} & \textbf{49.46} & \textbf{0.2397} \\
\bottomrule
\end{tabular}
\end{table}

\clearpage
\section{Experimental details for Feynman-Kac steering} \label{app:fk_details}

This section provides additional details on the Feynman-Kac steering experiments in \Cref{exp:feynman_kac}. All results use \ours{}-M and 200 SDE integration steps. For the MLIPs, we use \texttt{eSEN-ODAC25-Full} for eSEN, \texttt{uma-s-1p2} for UMA, \texttt{orb-v3-conservative-inf-omat} for Orb, and SevenNet-Omni with task \texttt{odac23}.

\subsection{Metric definitions}
We define three metrics: \texttt{MOFChecker} validity, structure validity, and formation energy error.
\begin{enumerate}
    \item \textbf{\texttt{MOFChecker} validity \citep{jablonka2023mofchecker}} (\%): an MOF is valid if it passes all \texttt{MOFChecker} checks, which include (but are not limited to) the presence of key elements (e.g., C/H and a metal), no atomic overlaps, chemically reasonable coordination environments (e.g., around C and N), sufficient porosity, no unphysically large charges, and no isolated molecules.
    \item \textbf{Structure validity} (\%): a predicted crystal structure is considered valid if (i) all pairwise interatomic distances are at least $0.5$~\AA, and (ii) the unit-cell volume is at least $0.1$~\AA$^3$.
    \item \textbf{Formation energy error} (eV/atom): the difference between the formation energy of the predicted structure and that of the ground-truth structure, measuring predicted stability relative to the ground truth (negative values indicate a more stable prediction):
    \begin{equation}
        \Delta E_f
        \;:=\;
        \Delta E_f(\hat{\mathcal{S}}) - \Delta E_f(\mathcal{S}).
    \end{equation}
    
\end{enumerate}




\subsection{Comparison with best-of-N reranking} \label{app:fk_best_of_n}

We compare FK steering with best-of-N reranking using $K$ particles or independent samples, respectively. Best-of-N selects the lowest-energy unsteered sample under \texttt{uma-s-1p2}; FK uses the same MLIP for steering, and both methods are evaluated with \texttt{eSEN-ODAC25-Full}. As shown in \Cref{tab:fk_best_of_n}, FK improves formation energy and structural validity over best-of-N. It is more effective for larger MOFs ($N\geq300$), with structural-validity gains 2.33--2.53 percentage points larger and reductions in $\Delta E_f$ 3.1--10.1 times greater than on the full evaluation set.

\begin{table}[ht]
\centering
\small
\setlength{\tabcolsep}{4pt}
\caption{\textbf{FK steering versus best-of-N at matched sample counts.}
Each entry reports best-of-N / FK, with the better value in bold.
$K$ denotes the number of samples or particles.
Results cover 1{,}000 randomly sampled BW test structures.}
\label{tab:fk_best_of_n}
\begin{tabular}{llccc}
\toprule
Data & $K$ & MOF validity (\%) $\uparrow$ & Structural validity (\%) $\uparrow$ & $\Delta E_f$ (eV/atom) $\downarrow$ \\
\midrule
All samples & 8 & 43.10 / \textbf{44.30} & 93.50 / \textbf{94.40} & 0.02128 / \textbf{0.01924} \\
 & 16 & \textbf{45.80} / 45.40 & 94.90 / \textbf{95.60} & 0.00989 / \textbf{0.00560} \\
\midrule
$N\geq300$ & 8 & 9.68 / 9.68 & 80.65 / \textbf{83.87} & 0.11992 / \textbf{0.09925} \\
 & 16 & 6.45 / \textbf{9.68} & 83.87 / \textbf{87.10} & 0.09898 / \textbf{0.08562} \\
\bottomrule
\end{tabular}
\end{table}

\subsection{Compute overhead of FK steering}

This section quantifies the additional inference-time cost of FK steering. We report inference-only compute for 1{,}000 BW test structures using the same checkpoint, 200 sampling steps, one sample per structure, and 8 NVIDIA H200 GPUs. As shown in \Cref{tab:fk_compute}, unsteered sampling takes 87 s, whereas FK steering increases runtime to 418 s at $K=4$ and 2{,}682 s at $K=32$, corresponding to 0.93-5.96 GPU-hours and a 4.80-30.83$\times$ slowdown.

This overhead is expected because FK steering maintains $K$ particles per sample and periodically evaluates an external MLIP potential for resampling. Since batching was kept largely fixed, the overhead could likely be reduced by adjusting batch sizes to fit the memory.

\begin{table}[h]
\centering
\caption{\textbf{Inference-time compute for FK steering on BW.} Results cover 1{,}000 test structures with the same checkpoint, 200 sampling steps, one sample per structure, and 8 NVIDIA H200 GPUs.}
\label{tab:fk_compute}
\begin{tabular}{lrrrrr}
\toprule
\textbf{Setting} & \textbf{$K$} & \textbf{Wall-clock (s)} & \textbf{GPU-hours} & \textbf{Samples/sec} & \textbf{Slowdown} \\
\midrule
No steering & -- & 87 & 0.19 & 11.494 & $1.00\times$ \\
\midrule
\multirow{4}{*}{FK steering}
 & 4 & 418 & 0.93 & 2.392 & $4.80\times$ \\
 & 8 & 786 & 1.75 & 1.272 & $9.03\times$ \\
 & 16 & 1511 & 3.36 & 0.662 & $17.37\times$ \\
 & 32 & 2682 & 5.96 & 0.373 & $30.83\times$ \\
\bottomrule
\end{tabular}
\end{table}

\subsection{Sensitivity analysis}

This section summarizes the hyperparameter settings used in the Feynman-Kac steering sensitivity analysis (\Cref{exp:feynman_kac}). To isolate the effect of each hyperparameter, we fix a default configuration and vary one parameter at a time while keeping all others unchanged. The default settings are listed in \Cref{tab:fk_default_config}. For completeness, we report the numerical results underlying \Cref{fig:fk_hyperparam} in \Cref{tab:steering_results}.

\begin{table}[h]
    \centering
    \caption{\textbf{Default configurations for Feynman-Kac steering sensitivity analysis.} The hyperparameters we consider are the number of particles ($K$), steering strength ($\lambda$), resampling interval ($\Delta \tau$), steering start time $t_{\text{start}}$, and the potential \citep{hartman2025controllable}.}
    \label{tab:fk_default_config}
    \begin{tabular}{l c}
        \toprule
        \textbf{Parameter} & \textbf{Value} \\
        \midrule
        $K$ & 8  \\
        $\lambda$ & 2.0  \\
        $\Delta \tau$ & 5 \\
        $t_{\text{start}}$ & 0.8 \\
        Potential & Immediate \\
        \bottomrule
    \end{tabular}
\end{table}

To evaluate the robustness of the method, we conducted ablation studies by varying individual hyperparameters. Table~\ref{tab:sensitivity_variants} summarizes the specific values tested for each parameter.

\begin{table}[h]
    \centering
    \caption{\textbf{Hyperparameter variations for sensitivity analysis.} For each experiment, one parameter was varied according to the values below, while all others remained at their default values.}
    \label{tab:sensitivity_variants}
    \begin{tabular}{l l}
        \toprule
        \textbf{Parameter} & \textbf{Varied Values} \\
        \midrule
        $K$ & 4, 16 \\
        $\lambda$ & 1.0, 5.0, 10.0 \\
        $\Delta \tau$ & 2, 5, 10 \\
        $t_{\text{start}}$ & 0, 0.20, 0.40, 0.60, 0.80 \\
        Potential & Difference, Max, Sum \\
        \bottomrule
    \end{tabular}
\end{table}

\begin{table}[h]
\centering
\caption{Numerical results for the FK steering sensitivity analysis (\Cref{fig:fk_hyperparam}).}
\label{tab:steering_results}
\begingroup
\setlength{\tabcolsep}{4pt}
\begin{tabular}{llccc}
\toprule
\textbf{Parameter} & \textbf{Value} & \textbf{\texttt{MOFChecker} validity} (\%) & \textbf{Structure validity} (\%) & \textbf{$\Delta E_f$} (eV/atom) \\
\midrule
\multirow{4}{*}{$K$}
 & 4 & 40.67 & 91.20 & 0.0548 \\
 & 8 & 44.02 & 94.00 & 0.0333 \\
 & 16 & 43.98 & 94.90 & 0.0173 \\
 & 32 & 47.27 & 95.50 & 0.0080 \\
\midrule
\multirow{4}{*}{$\lambda$}
 & 1.0 & 43.93 & 92.70 & 0.0425 \\
 & 2.0 & 44.02 & 94.00 & 0.0333 \\
 & 5.0 & 44.97 & 94.30 & 0.0286 \\
 & 10.0 & 43.89 & 94.60 & 0.0266 \\
\midrule
\multirow{4}{*}{$\Delta \tau$}
 & 2 & 42.58 & 93.10 & 0.0369 \\
 & 3 & 41.78 & 92.40 & 0.0346 \\
 & 5 & 44.02 & 94.00 & 0.0333 \\
 & 10 & 46.76 & 93.80 & 0.0292 \\
\midrule
\multirow{5}{*}{$t_{\text{start}}$}
 & 0.0 & 40.00 & 88.20 & 0.0927 \\
 & 0.2 & 42.52 & 89.80 & 0.0664 \\
 & 0.4 & 41.52 & 91.20 & 0.0519 \\
 & 0.6 & 42.65 & 91.90 & 0.0415 \\
 & 0.8 & 44.02 & 94.00 & 0.0333 \\
\midrule
\multirow{4}{*}{Potential}
 & Immediate & 44.02 & 94.00 & 0.0333 \\
 & Difference & 42.95 & 89.30 & 0.0797 \\
 & Max & 43.91 & 94.00 & 0.0337 \\
 & Sum & 44.15 & 93.70 & 0.0341 \\
\bottomrule
\end{tabular}
\endgroup
\end{table}

\clearpage
\section{MLIP-guided FK steering for inorganic crystal structure prediction}
\label{app:inorganic_csp_steering}

In principle, MLIP-guided FK steering applies to any flow matching model whose generative ODE can be converted to an SDE. To demonstrate generalization of our method beyond MOFs, we apply MLIP-guided FK steering to OMatG~\citep{hollmeropen}, an inorganic CSP model that operates with an SDE formulation. We use the released \texttt{Linear-SDE-Gamma} checkpoints for MP-20 and MPTS-52 with the default 310 integration steps, and use UMA (\texttt{uma-s-1p2}) as the guidance MLIP. FK hyperparameters match our main experiments: 16 particles, $\lambda=2.0$, immediate potential, resampling interval of 5, and start time $t=0.8$.

\Cref{tab:omatg_steering_results} shows that MLIP-guided FK steering consistently improves both metrics: structure validity increases by $+0.68\%$ (MP-20) and $+8.81\%$ (MPTS-52), and $\Delta E_f$ decreases by $72.8\%$ and $68.1\%$, respectively. These results suggest that MLIP-guided FK steering generalizes across diverse generative architectures and material types.

\begin{table}[h]
    \centering
    \small
    \caption{\textbf{FK steering for OMatG.}
    Steering OMatG~\citep{hollmeropen} with UMA~\citep{wood2025family} improves both structure validity and formation energy error on MP-20 and MPTS-52.}
    \label{tab:omatg_steering_results}
    \setlength{\tabcolsep}{8pt}
    \begin{tabular}{l l r r}
        \toprule
        \textbf{Dataset} & \textbf{Method} & \textbf{Validity (\%)} $\uparrow$ & \boldmath$\Delta E_f$ \textbf{(eV/atom)} $\downarrow$ \\
        \midrule
        \multirow{2}{*}{MP-20}
            & OMatG      & 99.26 & 1.4975 \\
            & OMatG + FK & 99.93 & 0.4074 \\
        \midrule
        \multirow{2}{*}{MPTS-52}
            & OMatG      & 91.37 & 4.2787 \\
            & OMatG + FK & 99.42 & 1.3652 \\
        \bottomrule
    \end{tabular}
\end{table}
\section{Experimental details for MOF-adsorbate structure prediction} \label{app:mof_ads_details}

\subsection{Baseline methods and implementation details}

\textbf{GCMC implementation.} For MOF atoms, we use a mixture of Dreiding~\citep{mayo1990dreiding} and UFF~\citep{rappe1992uff}; for adsorbates, we use TraPPE for \ce{CO2}, \ce{O2}, and \ce{N2}~\citep{potoff2001vapor, zhang2006direct}, and TIP4P-ew for \ce{H2O}~\citep{horn2005characterization}. Electrostatics use DDEC6 charges~\citep{manz2010chemically} predicted by \texttt{PACMAN}~\citep{zhao2024pacman} for MOF atoms and are computed with Ewald summation using a force tolerance of $10^{-6}$~kcal/mol/\angstrom; all pairwise interactions use a 12.8~\angstrom\ cutoff, and simulations are run at 298~K. During Monte Carlo sampling, translation, rotation, and reinsertion moves are attempted with equal probability. For each adsorbate type and loading, we run 10{,}000 initialization cycles followed by 10{,}000 production cycles and take the final production configuration as the representative MOF-adsorbate structure.

\textbf{Atomic overlap criterion for RS.} For RS, we reject insertions that create steric clashes, defined as any MOF-adsorbate or inter-adsorbate atom pair under periodic boundary conditions with minimum-image distance $d_{ij} \le 0.8\,(r_i + r_j)$, where $r_i$ and $r_j$ are ASE covalent radii~\citep{larsen2017atomic}.

\textbf{Learning-based baselines for ODAC25.}
Because our work is the first to model full MOF-adsorbate systems, there is no direct learning-based baseline for this task. We nevertheless attempted to adapt MOFFlow-2 to ODAC25, but found that it fails due to the complexity of experimental MOFs. ODAC25 contains 887 metal-node types with 1,384,427 variants, including oxo, hydroxo, water, and other bound species, whereas the hypothetical BW dataset contains only 9 metal-node types. Since these bound atoms vary across metal nodes, fixed-template models such as MOFFlow-2, MOFFlow, and MOF-BFN cannot generate or correct their internal all-atom geometry. Accordingly, the adapted MOFFlow-2 model achieves only 0.244\% 10-sample coverage on ODAC25, further motivating all-atom generative modeling for realistic MOF-adsorbate structures.

\subsection{Diversity of adsorption configurations}
\label{app:adsorption_diversity}

Coverage measures the recovery of reference adsorption configurations, whereas diversity quantifies variation among generated samples. Adapting the clustering-based metric of \citet{ram2026diffusion}, we count $C$ distinct clash-free clusters among $K=10$ configurations and define diversity as $100C/K\%$. Repeated identical clash-free predictions yield 10\% diversity, while entirely distinct clash-free predictions yield 100\%. We represent nearby framework environments using normalized periodic SOAP descriptors~\citep{bartok2013representing} within 6~\AA{} of each adsorbate atom, followed by complete-linkage clustering at a distance threshold of 0.10 to distinguish adsorption sites and orientations. We use the atomic-overlap criterion described above and average within each MOF--adsorbate system, then across systems.

\Cref{tab:adsorption_diversity} shows that \ours{} achieves substantially higher diversity than the deterministic reference, indicating that its predictions do not collapse to a single configuration. However, higher diversity does not necessarily imply greater practical effectiveness, since distinct but irrelevant configurations can increase this metric. For example, RS achieves nearly maximal diversity but the lowest coverage: 19.27\%, compared with 21.06\% for GCMC and 23.11\% for \ours{}. Moreover, GCMC and RS select configurations from substantially larger candidate pools. Together, diversity and coverage indicate that \ours{} generates varied configurations while recovering more reference-supported adsorption modes.

\begin{table}[ht]
\centering
\caption{\textbf{Diversity of adsorption configurations on ODAC25.} Results use 10 samples per input. The deterministic reference is analytical: repeated identical clash-free predictions produce one cluster.}
\label{tab:adsorption_diversity}
\begin{tabular}{lc}
\toprule
Method & Diversity (\%) \\
\midrule
Deterministic & 10.00 \\
\ours{} & 78.51 \\
GCMC & 89.31 \\
RS & 99.14 \\
\bottomrule
\end{tabular}
\end{table}

\section{Broader impacts} \label{app:broader_impacts}

Our work may help accelerate screening of MOFs for applications such as direct air capture by reducing the cost of evaluating candidate materials. However, the method could also be misused to screen materials for harmful purposes, and incorrect predictions could waste computational or experimental resources if used without validation. We therefore view \ours{} as a research tool for prioritization rather than a substitute for physical verification; its predictions should be followed by experiments, especially on large or out-of-distribution systems where reliability is lower.

\end{document}